\documentclass[article]{aa}
\pdfoutput=1
\usepackage{natbib,twoopt}
\usepackage[colorlinks,linkcolor=blue,urlcolor=blue,citecolor=black]{hyperref} 
\bibpunct{(}{)}{;}{a}{}{,}             
\makeatletter
  \newcommandtwoopt{\citeads}[3][][]{\href{http://ui.adsabs.harvard.edu/abs/#3}%
    {\def\hyper@linkstart##1##2{}%
     \let\hyper@linkend\@empty\citealp[#1][#2]{#3}}}
  \newcommandtwoopt{\citepads}[3][][]{\href{http://ui.adsabs.harvard.edu/abs/#3}%
    {\def\hyper@linkstart##1##2{}%
     \let\hyper@linkend\@empty\citep[#1][#2]{#3}}}
  \newcommandtwoopt{\citetads}[3][][]{\href{http://ui.adsabs.harvard.edu/abs/#3}%
    {\def\hyper@linkstart##1##2{}%
     \let\hyper@linkend\@empty\citet[#1][#2]{#3}}}
   \newcommandtwoopt{\citeauthorads}[3][][]%
    {\href{http://ui.adsabs.harvard.edu/abs/#3}
    {\def\hyper@linkstart##1##2{}%
     \let\hyper@linkend\@empty\citeauthor[#1][#2]{#3}}}
  \newcommandtwoopt{\citeyearads}[3][][]%
    {\href{http://ui.adsabs.harvard.edu/abs/#3}
    {\def\hyper@linkstart##1##2{}%
     \let\hyper@linkend\@empty\citeyear[#1][#2]{#3}}}
  \renewcommand*\aa@pageof{, page \thepage{} of \pageref*{LastPage}} 
\makeatother
\usepackage[varg]{txfonts}
\usepackage{pdflscape} 
\usepackage{booktabs}
\usepackage{csvsimple}
\usepackage{pgfplotstable}
\usepackage{array}
\usepackage{amsmath}
\usepackage{amsfonts}
\usepackage{dsfont}
\usepackage{amsxtra}
\usepackage{hyperref}
\usepackage{amssymb}
\usepackage{upgreek}
\usepackage{comment}

\usepackage{multirow}
\usepackage{url}

\usepackage{graphicx,epsfig}

\usepackage{xspace} 

\usepackage{lscape}

\usepackage{subfig}
\usepackage{xcolor}
\usepackage{lineno}
\begin{document}
%
%
%
%
\title{The origin of early-type runaway stars from open clusters }

   \author{A. Bhat \thanks{\email{aakash.bhat@fau.de}}
          \inst{1}
          \and
          A. Irrgang \thanks{\email{andreas.irrgang@fau.de}}
          \inst{1}
	\and
	U.Heber\thanks{\email{ulrich.heber@sternwarte.uni-erlangen.de}}
	\inst{1}
          }

   \institute{
             Dr.Karl-Remeis Observatory\& ECAP,
             Astronomical Institute, Friedrich-Alexander University Erlangen-N\"urnberg, 
             Sternwartstr.7, 96049, Bamberg, Germany
             }

\abstract
{
Runaway stars are ejected from their place of birth in the Galactic disk, with some young B-type runaways found several tens of kiloparsecs from the plane traveling at speeds beyond the escape velocity, which calls for violent ejection processes. Young open clusters are a likely place of origin, and ejection may be either through N-body interactions or in binary supernova explosions. The most energetic events may require dynamical interaction with massive black holes. 
The excellent quality of \textit{Gaia} astrometry opens up the path to study the kinematics of young runaway stars to such a high precision that the place of origin in open stellar clusters can be identified uniquely even when the star is a few kiloparsecs away. 
We developed an efficient minimization method to calculate whether two or more objects may come from the same place, which we tested against samples of Orion runaways. Our fitting procedure was then used to calculate trajectories for known runaway stars, which have previously been studied from \textit{Hipparcos} astrometry as well as known open clusters. For runaways in our sample we used \textit{Gaia} data and updated radial velocities, and found that only half of the sample could be classified as runaways. The other half of the sample moves so slowly ($<$30 km\,s$^{-1}$) that they have to be considered as walkaway stars. Most of the latter stars turned out to be binaries. We identified parent clusters for runaways based on their trajectories. We then used cluster age and flight time of the stars to investigate whether the ejection was likely due to a binary supernova or due to a dynamical ejection. 

In particular we show that the classical runaways AE Aurigae and $\mu$ Columbae might not have originated together, with $\mu$ Columbae having an earlier ejection from Collinder 69, a cluster near the ONC. 
The second sample investigated comprises a set of distant runaway B stars in the halo which have been studied carefully by quantitative spectral analyses. 
We are able to identify candidate parent clusters for at least four stars including the hyper-runaway candidate HIP 60350. The ejection events had to be very violent, ejecting stars at velocities as large as 150 to 400 km\,s$^{-1}$.

}

{

}
{
}
{

}
{}


\maketitle

\section{Introduction}
Decades ago, it was recognized that there are young stars present outside the spiral arms. These stars seemed to have left their place of birth and were moving away from the Galactic disk. \citetads{runawaysblaauw} studied the proper motions of two such stars, AE Aurigae and $\mu$ Columbae, and noticed that they were moving away from each other in opposite directions, with the direction of movement pointing away from the Orion nebula. These two stars formed a part of the larger survey of such stars by \citetads{runawaysblaauw2}, where the name "runaways" was used 
if their  peculiar velocities were greater than 40 km\,s$^{-1}$. \citetads{runawaysblaauw2} explained that such stars had been part of a binary system, and when one of the stars underwent a supernova, the other star was released with a velocity close to its original orbital velocity; we refer to this scenario as the binary supernova scenario (BSS). Many runaway stars have been discovered (e.g., \citeads{1986ApJS...61..419G},
\citeads{1997fbs..conf...87H}, 
\citeads{2001A&A...369..530T},
\citeads{2004MNRAS.349..821L},
\citeads{2008A&A...483L..21H},
\citeads{2015MNRAS.448.319six-dimensional},
\citeads{2019A&A...624A..66R},
\citeads{2021A&A...645A.108R}) since then.

An alternative scenario of dynamical ejection from stellar clusters was proposed (\citeads{1967BOTT....4...86P},
\citeads{fujizwart},
\citeads{originrun}). Accordingly, stars can be ejected from a stellar cluster during its initial dynamical relaxation. This scenario, known as the dynamical ejection scenario (DES) also explains the expansion of 
compact and loose stellar clusters (\citeads{2008MNRAS.389..223B},\citeads{2013A&A...559A..38P},
\citeads{2019MNRAS.483.4999F},
).

Regardless of the formation mechanism, most of the stars discovered so far are massive stars of spectral type O and B and can be divided into two types of young runaway stars. Young massive stars found in the halo which have already left the disk  of the Galaxy \citepads{2011MNRAS.411.2596S}, and stars which seem to be traveling away from stellar clusters at space velocities of ~30 km\,s$^{-1}$ \citepads{1986ApJS...61..419G} or more, but have yet to leave the disk.
 This lower velocity limit 
is somewhat arbitrary, but it separates the runaways from the class of walkaway stars, usually defined simply as being unbound from their parent associations (\citeads{de_Mink_2014},
\citeads{2019A&A...624A..66R}). Walkaway stars are similar to runaways in terms of their origin, but they are more abundant due to their lower velocities \citepads{2005A&A...437..247D}.
Since the average velocity dispersion in such associations has been found to be between 3-8 km\,s$^{-1}$ (\citeads{2009MNRAS.400..518M}, 
\citeads{2017MNRAS.472.3887M}), 
a strict threshold on the minimum velocity of walkaway stars has not been set. However, previous studies have used a lower value of 5 km\,s$^{-1}$ \citepads{2011MNRAS.414.3501E} to account for all walkaways, or 10 km\,s$^{-1}$ to take the maximum velocity dispersion in all associations into account \citepads{2020MNRAS.495.3104S}.

Although studied for decades, the origins of most runaways are still not known. Nevertheless, interest in kinematically peculiar stars, especially with the most extreme velocities, has never been lost. \citetads{hills} had theorized that stars with velocities above 1000 km\,s$^{-1}$ might also exist. A supermassive black hole at the center of the Galaxy could tidally disrupt a tight binary system and release one component at a velocity up to a few thousands km\,s$^{-1}$, which would be unbound to the Galaxy. However, it took until 2005, when hypervelocity stars that are unbound to the Galaxy were discovered 
(\citeads{brown1},
\citeads{us708},
\citeads{hypervelbamberg}). The \textit{Gaia}  mission \citepads{Gaia2} provided proper motions which, for the first time, allowed us to study the candidates in the full six-dimensional phase space. Indeed, SS5-HVS1 was discovered as an A-type star traveling at a rest frame  velocity of 1755 km\,s$^{-1}$ \citepads{10.1093/mnras/stz3081}. \textit{Gaia}-based results for other B-type candidates indicated that many of them do not originate from the Galactic center but from the Galactic disk \citepads{2018A&A...620A..48I}. 
Despite the high end of the velocity spectrum receiving much attention, studies have continued to show that most stars do not come from the center of the Galaxy \citepads{andreasnew}. The suspected source of origin in this case are star clusters. After birth a few of the stars are ejected either dynamically through interaction with other stars, or released when its companion in a binary or triple system undergoes a supernova.
The aim of this paper is to identify the parent clusters of known massive runaway stars (mostly of spectral type B). In Sect. \ref{sect:trajectories} our calculations of trajectories in a Galactic potential is described. Sect. \ref{sect:fitting} describes a new fitting method. The Orion nuclear cluster is used as a testbed for the fitting method (Sect. \ref{sect:onc}).  In Sect. \ref{sect:hoogerwerf} we revisit the sample of cluster runaways investigated by \citetads{hoog}. 
In Sect. \ref{sect:high_vel}  and \ref{sect:hip60350} we apply our new method to B type runaways in the Galactic halo and summarize the results and conclude in Sect. \ref{sect:summary}.

\section{Trajectories}\label{sect:trajectories}
To provide evidence of origin of stars in clusters, we need to trace back stars and clusters in time. There are two main requirements to solve this problem. Firstly, the Galactic potential of the Milky Way needs to be modeled, which requires an understanding of the structure of the Milky Way. 
Secondly, the trajectory of a test particle (in this case a star or a cluster) under the influence of a potential needs to be calculated numerically.

\citetads{allen1} proposed an analytic model for the Milky Way's gravitational potential. Their model consists of three axis-symmetric components: A spherical central mass or bulge, a disk, and a spherical halo. 
The model has since been used to calculate and pinpoint the orbits of hypervelocity stars in many studies, for example by \citetads{1992AN....313...69O}. 
The parameters of this model have been updated by \citepads{andreassmass} to match recent observations \citepads[e.g.][]{andreasstar}.
We use this model \citepads[model I of][]{andreassmass} for all computations of trajectories {\bf using the most recent version of 
the orbit integrator by \citetads{andreasnew} as well as \texttt{galpy} \citepads{bovy}.}
The second part of the analysis is to find out if two objects come from the same place at the same time, which has so far required Monte Carlo simulations. Therefore, the kinematic parameters (distances, positions, proper motions, and radial velocities) of both of the objects are first acquired. 

Typically, a Monte Carlo simulation is run which randomizes the value of the parameters within their uncertainties. For each run, the trajectory of both objects in the Galactic potential is computed, and the minimum distance between the two is calculated for all possible combinations of trajectories. This process requires a computation of millions of trajectories and the acquired histograms provide insights into whether the two stars (or a star and a cluster) were close enough to be considered as having a similar origin or not. In this paper, we shall employ a new method of finding sites of origin for runaway stars. Our method is based on a minimization procedure, and we shall use it on candidate lists of stars provided in Sect. \ref{sect:hoogerwerf} to \ref{sect:hip60350}.

\section{Fitting procedure}\label{sect:fitting}

\citetads{hoog} provide mathematical evidence for why it is not possible for the distribution of the closest distance of encounter between two objects to peak at zero, even if the two objects come from the same space-time point. The reason is that in three-dimensional space the errors on the observed parameters shift the observed mean of the closest possible distance. For two objects we have at least eight parameters (the two parallaxes, the four proper motion components, and the two radial velocities) with nonzero standard deviations. To claim that two objects come from the same place at some time, they relied primarily on simulations of trajectories of the two stars in a Galactic potential. These simulations have since been used by other authors (\citeads{andreasstar}; \citeads{2020MNRAS.498..899N}; \citeads{2021AstL...47..224B}) to search for the origins of runaways and hypervelocity stars. However, these simulations have two shortcomings. Firstly, they are time consuming. For example, to model only a few thousand combinations of trajectories imposing an encounter distance of less than 10 pc, at least a million combinations of trajectories need to be calculated. For instance, for $\zeta$ Oph and PSR J1932+1059, only 4214 trajectories out of three million trajectories simulated by \citetads{hoog} came closer than ten pc. Secondly, due to time and memory constraints, simulations are highly inefficient for shortlisting candidate lists of potential runaway stars. This is especially pertinent for candidates which are more than a kpc away, and might require computations of trajectories up to hundred Myrs.
 
In the following we introduce a way to more efficiently calculate whether two or more objects can be said to come from the same place. Our method is based on a fitting procedure which fits the trajectories of the objects by  minimizing the distance between them using a $\chi^2$ method.  
This technique is computationally much less costly than Monte Carlo simulations.

The fitting procedure allows us to find the minimum distance possible between the trajectories of two or more objects. The trajectories are calculated using a gravitational potential, as defined in section \ref{sect:trajectories}. The first trajectories use the mean values of the kinematic parameters. We then calculate the distances between all points in these set of trajectories, which allows us to find the minimum distance for the set of trajectories. This distance is then required to be minimized in order to find trajectories which have the least minimum possible distance. The complete algorithm then followed the following steps:

1. The trajectories for two or more objects were computed using the mean values of kinematic parameters (for n objects these are 6*n mean values). 2. The distance between the objects was calculated for all time steps till the maximum allowed time. The minimum of this distance provides the first guess for the minimization procedure. 3. This distance was kept as the function to be minimized. This was followed by minimization of the distance function, wherein the kinematic parameters were allowed to vary within their uncertainties (for n objects these were 4*n parameters, since the initial right ascension and declination were kept constant). 4. The minimization was stopped when a certain lowest distance was reached. Ideally, the distance for two point like sources may be close to zero, although the required threshold was kept higher.

A drawback of the $\chi^2$ distance minimization approach is that the trajectories may be statistically insignificant. For instance, it is possible to find trajectories which have parameter values 4-5$\sigma$ away from their means, simply due to the fact that these trajectories might provide the best fits. Therefore, to remedy this, we created another measure for our minimization procedure which took into account the significance of our trajectories by weighing the significance of the final trajectory in our distance function through the use of the statistical p-value. 

We calculate the average distance between 'N' objects at any time as a function $D_N$. This distance (given here in kiloparsec) is calculated using the Galactic potential model 1 of \citeads{andreassmass}. For more than 2 objects, we take the average distance between them. To add a measure of significance, we define the parameter $D_p$ as:
\begin{align}
 D_p=\text{wf}*(1-p), \label{eq.2}
\end{align}
The multiplicative factor wf is a weighting factor which allows us to control how much weight we want to allow this function to have. In our algorithm we use the p-value, given as $p$ here, which we calculate using the $\chi^2$ method. Since we have many parameters, the final p-value is calculated using all the parameters and serves as a metric for how close the trajectories are to the mean trajectories.

Finally, we minimize a function which takes into account both $D_{N}$ and $D_{p}$, defined as:
\begin{align}
D_M: \sqrt{D_ N}+ D_p \label{eq.1}
\end{align}

The final value of $D_M$ is therefore highly dependent on the choice of weighting factor. We show here that subject to the condition that $D_N$ be less than a predefined chosen value $\delta$, it is possible to choose a wf such that both $D_N$ and $D_p$ are almost equally weighted. The chosen value $\delta$ then defines a 'close encounter'. For our fitting procedure throughout this paper we choose $\delta$ as 1 pc \footnote{Since the closest distance may theoretically vary from 1 pc to 0, it has a greater spread than the p-values we are interested in, i.e., greater than 2 sigma. Therefore we chose to take the square root for a nonlinear mapping of p-values.} . 
Therefore, considering this value to be the maximum possible minimized value, we choose to have $D_N=0.1$ pc to validate that two or more objects come from the same place. We assume that the p-value should be roughly 0.9. For reference, a value $p>0.43$ for 8-parameters corresponds to all 8 parameter values being within 1$\sigma$ of their means. A value of 1.0 may be chosen, but is practically very difficult to achieve when performing any minimization. Therefore, for equal weighting we have from Eq.\ref{eq.1} and \ref{eq.2},
\begin{align}
\sqrt{10^{-4}} = 10^{-1}*\text{w.f}\\
\implies \text{w.f} = 10^{-1}
\end{align}

Hence, our procedure minimizes the quantity:
\begin{align}
D_M =\sqrt{D_N} + 0.1*(1-p)
\end{align}

This implies that step 4 in our procedure was changed to: Stop the minimization when the best fit is found for a distance lower than the threshold distance and a p-value higher than the chosen significance level. 

\noindent The adopted null hypothesis states that the two or more objects whose trajectories we are optimizing have an optimal trajectory which originates from the same space-time point. The significance level, a p-value below which would allow us to reject the null hypothesis, should therefore be decided before hand. This level, also referred to as $\alpha$, depends on the number of parameters we fit for our trajectories, as well as the method used to compute it. Therefore, common $\alpha$ values are not conceptually the correct ones for our work, since they are chosen based on 1 parameter Gaussian distributions. As can be seen in Table \ref{tab:significance}, the 8 parameter minimization can lead to very low p-values when compared to the 1 parameter solution. For our present paper, we chose the $2\sigma$ rejection limit for 8 parameters. Actually, we aim at the failure of rejecting the null hypothesis and, therefore, we regard any p-value above the $1\sigma$ rejection limit as support to the null hypothesis. This implies a stringent condition that there is a common origin for two objects where all parameters of the objects are within $1\sigma$ of their observed means. 

\begin{table}
 \begin{center}
\renewcommand{\arraystretch}{1.3}
\begin{tabular}{c c c  } 
\hline
\hline
Significance $\alpha$& P-val(1 Parameter)&P-val\\
 &(1 Parameters)  &  (8 Parameters)\\
\hline
$1\sigma$ & 0.31 &0.43\\ 
$2\sigma$&0.04&$9*10^{-5}$ \\ 
$3\sigma$&0.002&$1.9*10^{-12}$\\
\hline
\end{tabular}
\caption{P-values corresponding to different significance levels for 1 and 8 parameters using the $\chi^2$ distribution when all parameters are at the respective ends of the uncertainty.}
 \label{tab:significance}
\end{center}
\end{table}
\subsection{Extent of clusters}\label{cluster_extent}

\begin{figure}[htp]
\hspace*{-1cm}
\centering
\includegraphics[width=0.6\textwidth]{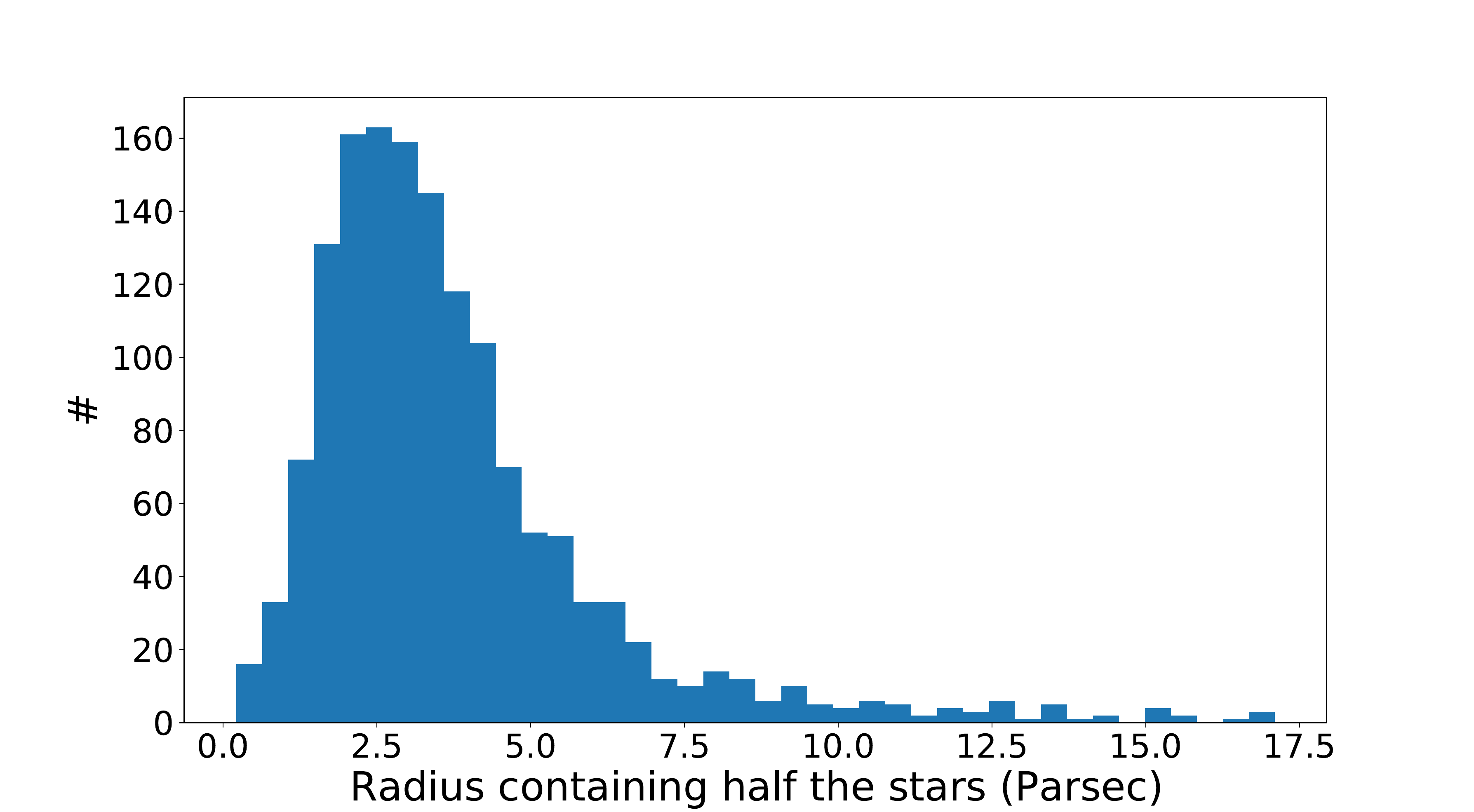}
\caption{Cluster extent}
\label{fig:clusterextent}
\end{figure}

 The minimum spatial distance $\delta$, considering the extent of open clusters on the sky, may be considered acceptable up to 10 pc for close encounters. However, we are more conservative by requiring that the minimum distance be less than 1 pc from the cluster center. Most of the runaway ejections are thought to happen here due to the density and compactness of the cluster cores. \citetads{DR2OC} provide a list of clusters with parallaxes and angular extent of the clusters. We used these to compute the radius of the clusters within which 50\% of the cluster members are located (see Table \ref{tab:extentofclusters}
 and Fig. \ref{fig:clusterextent}). The mean of the histogram is  $\approx$3.7 pc with a standard deviation of 2.4 pc. This is slightly higher than our chosen significance level and therefore, in case the minimization fails for a particular star/cluster or association, the cases are inspected individually allowing for somewhat larger distances for close encounters. For instance, the stars HIP 22061 and HIP 29678 both come close to the cluster Collinder 69, that is 7 pc from the center of the cluster, which is identical with the extent of this cluster. Hence, the two stars may have been ejected from the outskirts of the cluster rather than its central region. The rigid criterion also allows us to include small clusters.
 
\begin{table}
\begin{center}
\renewcommand{\arraystretch}{1.2}
\begin{tabular}{c c } 
\hline
\hline
Cluster& Extent (pc)\\
\hline
Cl Alessi 20&1.77\\
Cl Gulliver 9&8.51\\
Pozzo 1&3.8\\
NGC3532&4.52\\
$\eta$ Cha Association&0.62\\
Cl Alessi 13&2.47\\
\hline
\end{tabular}
\caption{Radius containing 50\% of the member stars of new clusters reported in this paper.}
 \label{tab:extentofclusters}
\end{center}
\end{table}
 
\noindent Table \ref{tab:extentofclusters} shows the radius for some of the new clusters we have studied in this paper. 

\subsection{Age of clusters and runaway stars}\label{time_character}
The high accuracy of \textit{Gaia} astrometry has allowed the CMDs of open clusters to be cleaned by improving cluster membership from precise common proper motions. Hence, the cluster proper motions are based on large number of cluster members (see Table \ref{tab:clusterparams}). Based on such cleaned CMDs ages were recently derived by isochrone fitting for many clusters \citepads[e.g.][]{2019A&A...623A.108B,Liu}. In many cases cluster ages are given in the literature with statistical uncertainties only, based on the favored isochrone models. The role of different  stellar models has been investigated in a few cases \citepads[e.g. for the well studied cluster Alessi~13,][]{2021A&A...654A.122G}.
While common origin of two or more objects can be constrained spatio-temporally by using the distance between them at a certain time, the characterization of the method of origin into BSS or DES requires further study. The DES method happens during the formative years of the clusters (\citeads{hoog};\citeads{fujizwart}). Therefore, for early type massive stars the cluster age and flight time should be comparable. On the other hand the BSS requires the evolution time of the primary star. This implies that the cluster should not be so old that it precedes the formation of the O- or B-type stars that were the primary, and the difference between the age of the cluster and the time of flight of a BSS runaway should be bounded. For this reason, a cluster may theoretically have a maximum of age of $\approx$150 Myrs \citepads{2013A&A...558A.103G}, which is roughly the age of a 4 M$_\odot$ B-type star. This age can be slightly greater, since BSS candidates can appear younger due to the rejuvenation by the primary. They are then known as blue stragglers, which have been predicted to exist as a result of collisions in open clusters \citepads{1999A&A...348..117P}, have been expected and found in ejected populations of runaway stars \citepads{2000ApJ...544..437P,hoog}, and have even been suggested to explain the place of origin of hypervelocity stars \citepads{2009ApJ...698.1330P,2010ApJ...719L..23B}. Theoretically, considering a minimum mass of nine solar masses required for a supernova, the approximate lifetime of 30 Myrs \citepads{2013A&A...558A.103G} should be considered for a BSS progenitor, and therefore the difference in the cluster age and the flight time of the runaway star should not exceed 30 Myrs. Due to the existence of blue stragglers and the fact that clusters ages might be inaccurate, this difference may appear greater in practice.

In the following we first look for clusters with a spatial coincidence with runaway stars in the past. We then use cluster ages to confirm or rule out common origins.
\subsection{A comparison of measurements}
\begin{figure*}[h]
\hspace{-0.5cm}
\includegraphics[width=1.0\textwidth]{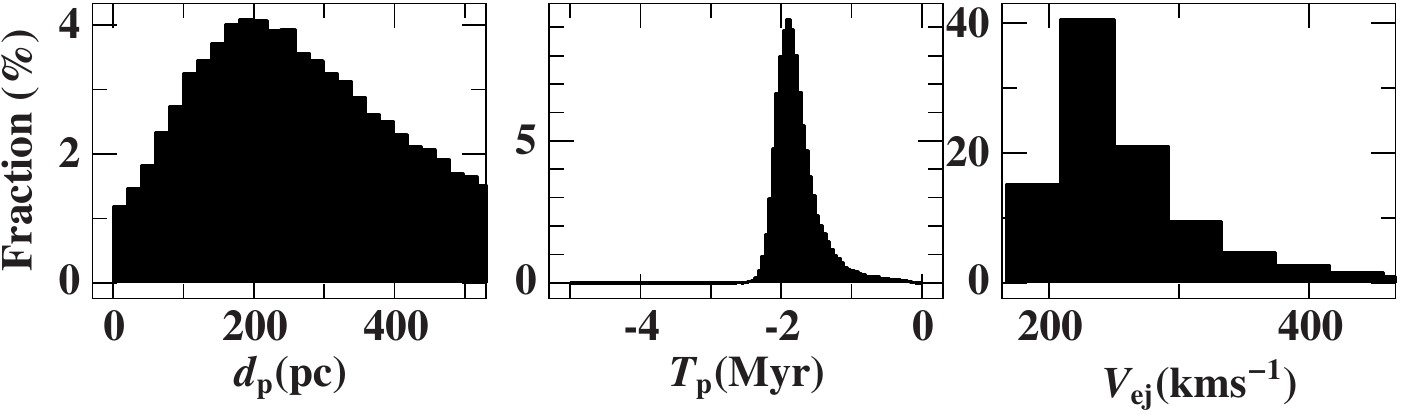}
\hspace*{-0.25cm}
\includegraphics[width=1.0\textwidth]{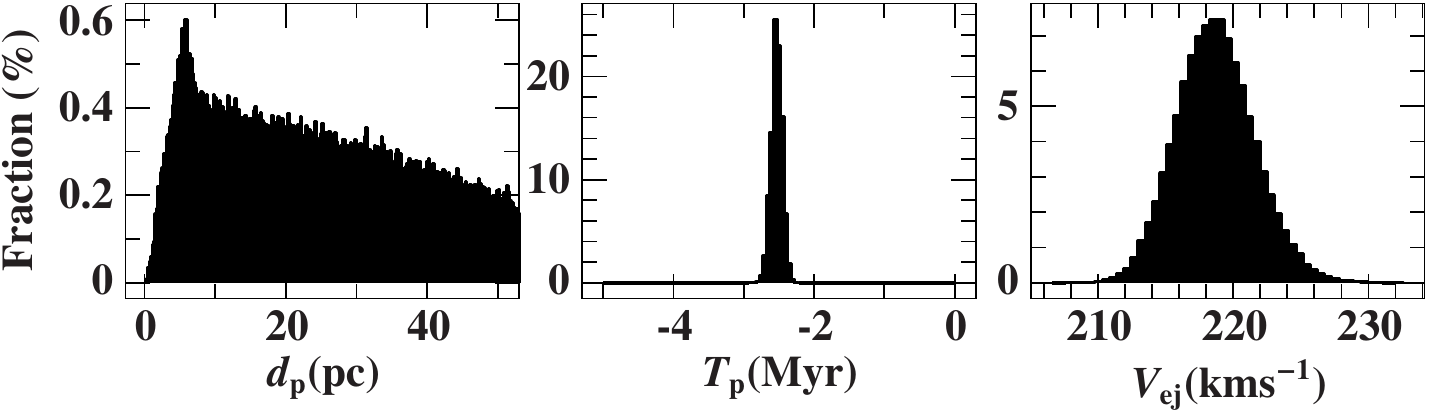}
\caption{Histograms for back-traced orbits of AE Aurigae and $\mu$ Col from Monte Carlo simulations. with \textit{Hipparcos} second reduction (top) and \textit{Gaia} EDR3 data (bottom). The left most panel shows the fraction of distances of closest encounters between the two stars. The middle panel is the histogram of the times for the closest encounters of the two stars, and the right most panel plots the relative velocity of the stars with respect to each other.}
\label{fig:hist}
\end{figure*}

To check whether the accuracy and precision of the new measurements is comparable to our distance thresholds, we ran 100000 simulations of orbits for AE Aurigae and $\mu$ Col, using Hipparcos second reduction and Gaia EDR3 data. The results shown in Fig. \ref{fig:hist} for EDR3 show most distances close to 10 pc (lower panel), which is more than an order of magnitude improvement compared with results from Hipparcos data (upper panel and is mainly a consequence of the factor 10 improvement of the proper motions (see Fig. \ref{fig:parallaxcomp}). Owing to this, we find that our $\chi^2$ fitting procedure, which is designed to target significant orbits with closest distances below 1 pc, may indeed be used with the present day measurements.
\subsection{AE Aur and $\mu$ Col}\label{aeaur}

The first tests done to validate our method were for the two well studied stars AE Aurigae and $\mu$ Columbae, which have been shown to have originated from the same space $\approx$2.5 Myrs ago. We relied on the same \textit{Hipparcos} astrometry as \citet{hoog} to carry out a first test. 
Using the minimization method above and allowing for a variation of each parameter up to $2\sigma$ uncertainty, allowed us to find the best fit times and the corresponding p-values. We found a p-value of 0.72 using \textit{Hipparcos} astrometry, suggesting a significant encounter between the two stars 2 Myrs ago. This p-value was even higher when the parameters were allowed to vary only till $1\sigma$ uncertainty, which is a much more stringent condition in this scenario. We found a p-value of 0.87, which implies that there exist trajectories for which AE Aurigae and $\mu$ Col originated at the same point, with all kinematic parameters close to their observed means.

As a test of our metric, we also ran the fitting but without the p-value term (wf was kept 0). For the $2\sigma$ case, the a posteriori calculated p-value of the minimized trajectory was 0.002, a drastic reduction from the value of 0.72 before. 
This allows us to conclude that the addition of the p-value term into the distance metric has two important consequences. Firstly, the explicit p-value term allows to find more statistically significant trajectories that were close to each other compared to an unweighted fitting procedure. Secondly, the introduction of the p-value provides us with a statistical quantity using which we may confirm or reject the null hypothesis.



\section{The Orion nuclear cluster -- a testbed}\label{sect:onc}


In order to carry out another test based on actual observations we applied our fitting procedure to other suggested runaways from 
the ONC which is close and young enough \citepads{2019ApJ...884....6M} to study its runaway population. It has been studied quite extensively and  \citetads{2020MNRAS.495.3104S} used \textit{Gaia}  DR2 measurements to search for runaways and walkaway candidates from the ONC. Their list of stars were based on 2-D velocity tracebacks where the ONC was kept stationary. In addition to the traceback, they also used isochrone measurements to shortlist candidates based on their age. A final addition was that of radial velocities which were taken from \textit{Gaia} DR2 \citepads{2019A&A...622A.205K}. Unlike the sample of massive early-type stars of \citetads{hoog}, the sample consists of late-type stars. Consequently, there were 10 stars reported to be three-dimensional runaway candidates using these radial velocities. In addition, we  supplemented the \textit{Gaia} DR2 \citepads{2018A&A...616A...2L} with EDR3 measurements \citepads{2021A&A...649A...2L} along with  \textit{Gaia}  DR2 radial velocities \citeads[as used by][]{2020MNRAS.495.3104S}. Similarly, \citetads{Farias_orion} provide a list of 16 stars of late spectral type also based on {\textit  Gaia}  DR2 astrometry with corresponding \textit{Gaia}  DR2 radial velocities. Their method also relies on multiple flags like age and young stellar object color and magnitude, 2-d closest distance of approach, and three-dimensional overlap. 
Table \ref{tab:schoetller} list p-values from our calculations for all these 26 stars based on both \textit{Gaia} DR2 and EDR3 astrometry, where the lower limit was fixed to D$_\mathrm{p}$=5 pc that is the extent of the ONC. We also report the time and minimized distances for these stars. Of particular note here is that the stars have significantly high p-values. This result is consistent both with our method finding the minimized distance, and with the method used by \citetads{2020MNRAS.495.3104S}. In their paper, the mean values of the stars are used to find the velocities in the ONC rest frame. Since this method relies on the mean values to shortlist runaway stars it is consistent with high p-values since the observed parameters tend to be closer to their observed means.

\begin{table*}[ht]
\begin{center}
\caption{ P values for the three-dimensional runaway samples of \citetads{2020MNRAS.495.3104S} and \citetads{Farias_orion}, thought to come from the ONC. }
\hspace*{-0.5 cm}
\renewcommand{\arraystretch}{1.3}
\begin{tabular}{ c c c c c c c c c  } 
\hline
\hline
Number& \textit{Gaia}  DR2 Source ID& P2& T$_f$&P3& T$_f$&D$_\textrm{p,2}$ & D$_\textrm{p,3}$\\
\hline
1&3216203177762381952& 0.993&-0.229 &  0.576&-0.231&$<1$pc&$<1$pc  \\
2&3012438796685305728 &0.993&-0.544& 0.942&-0.545&1.2 pc&1.6 pc\\
3&2986587942582891264 &0.786&-1.391& 0.988&-1.394&3.2 pc&4.0 pc\\
4&3017250019053914368 &0.961&-0.038& 0&-0.042&1.14 pc&0.9 pc\\
5&3122561556293863552 &0.987&-2.527& 0.989&-2.274&2.3 pc&2.6 pc\\
6&3017265515291765760 &0.999&-0.0009& 0.842&-0.0007&1.2 pc&7.5 pc\\
7&3222673430030590592 &0.983&-1.606& 0.970&-1.597&0.8 pc&1.2 pc\\
8&2988494014709554176 &0&-2.758& 0&-2.788 &$<<1$ pc&$<<1$ pc\\
9&3208970285334738944 &0.982&-0.524& 0.998&-0.537 &2.6 pc&2.5 pc\\
10&3015532208227085824 &0.974&-0.972& 0.939&-0.843 &2.0 pc&3.7 pc\\
\hline
\hline
11&2983790269606043648&0.86&-1.05&0.40&-1.05&1 pc&1 pc \\
12&2963542281945430400&0.94&-2.31&0.94&-2.38&86 pc&88 pc \\
13&3009308457018637824&0.41&-0.84&0.88&-0.84&$<<$ 1 pc&0.15 pc \\
14&3021115184676332288&0.98&-0.88&0.97&-0.87&0.2 pc&0.19 pc \\
15&3008883530134150016&0.28&-0.81&0.93&-1.19&0.7 pc&21 pc\\
16&2969823139038651008&0.90&-1.68&0.91&-1.78&40 pc&46 pc\\
17&2984725369883664384&0.96&-2.28&0.96&-2.31&83 pc&83 pc\\
18&3184037106827136128&0.19&-1.65&0.26&-1.62&$<<1$ pc&$<<1$ pc\\
19&3209424795953358720&0.95&-0.36&0.88&-1.26&3.6 pc&3.4 pc\\
20&3017364028971010432&1.0&-0.0009&0.94&-0.41&0.14 pc&1.6 pc\\
21&3017367391918532992&0.95&-0.96&0.95&-0.93&1.1 pc&0.2 pc\\
22&3017359871442791168&0.97&-0.15&0.97&-0.15&$<<$1 pc&$<<$1 pc\\
23&3017358978089804672&0.99&-0.27&0.86&-0.28&0.1 pc&0.2 pc\\
24&3017364544367271936&0.99&-$10^{-5} $&0.96&-0.0007&0.7 pc&0.7 pc\\
25&3209521037582290304&0.99&-0.1&0.99&-0.32&0.9 pc&1 pc\\
26&3017360554330360320&0.98&-0.15&0.93&-0.17&0.2 pc&0.2 pc\\
\hline
\end{tabular}
\tablefoot{P2 and P3 are the p-values using \textit{Gaia}  DR2 \citepads{2018A&A...616A...2L}  and EDR3 data \citepads{2021A&A...649A...2L} , respectively, 'T$_f$' is the corresponding time of flight in Myrs, and D2 and D3 are the minimum distances using DR2 and EDR3.}
 \label{tab:schoetller}
\end{center}
\end{table*}

The minimized times of flight we find are also within the same range as reported by the authors, and especially in the case of stars numbered 4, 6, and 24 the minimized times are much less than 1 Myrs, meaning that they might indeed be within the cluster. However, for the latter star the minimized distance with EDR3 is 7.5 pc, which may hint at this being a nearby star and not a star from the cluster. Furthermore, while most of the stars have minimized distances below or near the nominal radius of 2.5 pc, the stars 12, 16, and 17  had close approaches greater than 40 pc, implying that these stars might not have their origin in the ONC. For star 15, the 20 pc difference between DR2 and EDR3 may be explained by the different parallaxes of 3.13 and 3.27 in DR2 and EDR3 respectively.

\section{The Hoogerwerf et al. sample revisited}\label{sect:hoogerwerf}
 \citetads{hoog} in their seminal paper curated a list of 21 runaway stars based on \textit{Hipparcos} astrometry
 along with their hypothesized places of origins. These origin sites were either nearby associations or open clusters, using the original \textit{Hipparcos} catalog \citepads{1997ESASP1200.....E} for parallaxes and proper motions of the stars. 
 This catalog has been re-reduced by \citetads{2007ASSL..350.....V} and 
 more precise measurements have been obtained by the \textit{Gaia}  satellite \citepads {2016A&A...595A...1G}, in particular the data release 2 \citepads[DR2][]{Gaia2}) and the early data release 3 (EDR3, \citeads{gaia3}). We compile the astrometric data from the new \textit{Hipparcos} reduction
 as well as the two \textit{Gaia}  data releases in Table \ref{tab:hypervelparams}. As quality indicators we use the goodness of fit parameter F2 for the \textit{Hipparcos} data and the renormalized unit weight error (RUWE) for \textit{Gaia} , requiring |F2|<5 and RUWE<1.4, respectively. Furthermore, since the \textit{Gaia}  brightness limit saturates at 3.6 mag, we do not use \textit{Gaia}   data for stars brighter than this. Therefore, for the stars $\zeta$ Oph, $\zeta$ Pup, and $\iota$ Ori, we stick to \textit{Hipparcos} astrometry. The \textit{Hipparcos}
based distance has been reconfirmed by \citetads{2019MNRAS.484.5350H}. The RUWE parameters of \textit{Gaia}  EDR3 of two more stars, HIP 18614 and HIP 57669 exceed the limit and, therefore, their astrometry is taken from {\textit {Gaia} }  DR2. For the rest of the stars we stick to {\textit  Gaia}  EDR3. Finally, we add the star HIP 82286 (HD 151397) which was studied by Blauuw to this list as it is a disk runaway. The spectro-photometric distance and radial velocity for this star have been provided by \citetads{Markus:Thesis:2021}, and we use \textit{Gaia} EDR3 for the astrometry.
 
\begin{figure}[htp]
\centering
\includegraphics[width=0.5\textwidth]{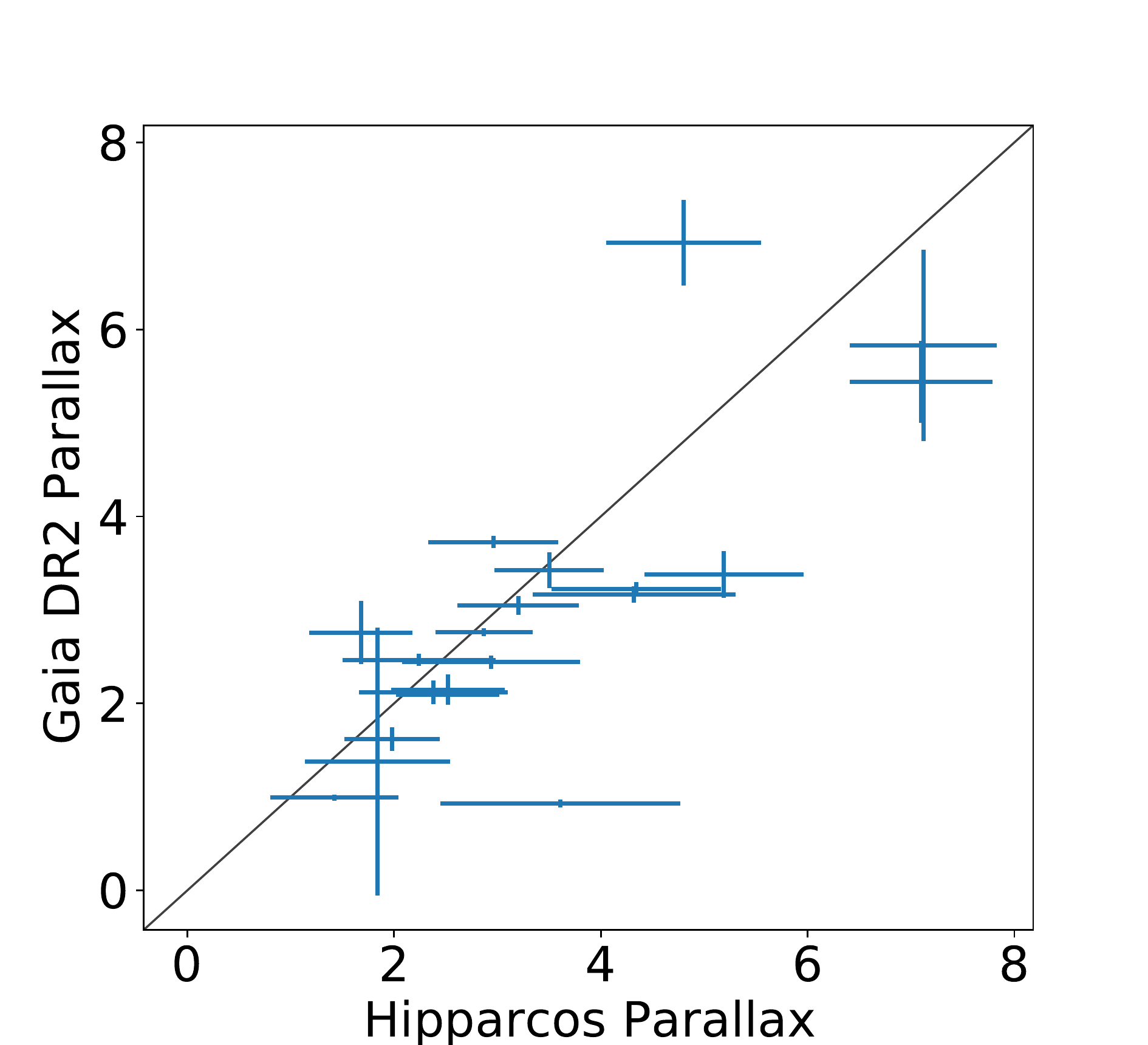}
\caption{Comparison of \textit{Hipparcos} parallax values used by \citetads{hoog} with {\textit  Gaia } DR2 values. %
}
\label{fig:parallaxcomp}
\end{figure}

 Fig. \ref{fig:parallaxcomp} compares \textit{Gaia} DR2 and \textit{Hipparcos} parallaxes for the runaway stars of \citetads{hoog}, where corresponding \textit{Gaia}  parallaxes are available. For most of these stars, \textit{Gaia} provides an order of magnitude better data, except for the brightest stars. 
 
\begin{figure}[htp]
\hspace{-0.5cm}
\centering
\includegraphics[width=0.5\textwidth]{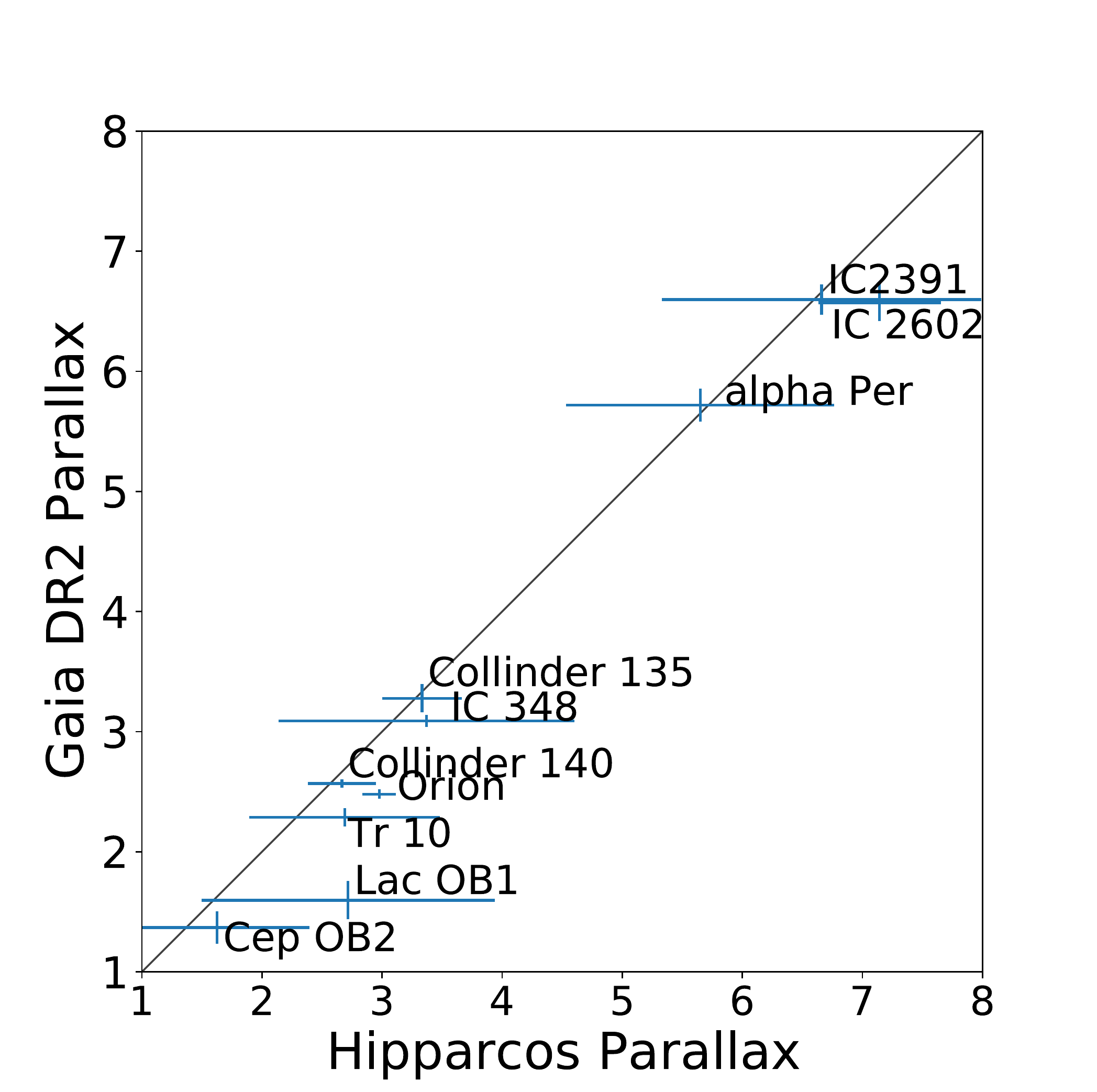}
\caption{Comparison of parallax values used by \citetads{hoog} with \textit{Gaia}  DR2 based values for clusters mentioned in \citeads{hoog}.}
\label{fig:parallaxcompclus}
\end{figure}

 Since {\textit  Gaia}'s catalog was released, hundreds of new clusters have been identified 
 (\citeads{DR2OC},
 \citeads{DR2OC2},
 \citeads{GaiaHR},
 \citeads{Liu}).
 Apart from new clusters, the parameters of known clusters have been updated to more precise values (see Table \ref{tab:clusterparams}), as can also be seen in Fig. \ref{fig:parallaxcompclus}. Therefore, it is timely to revisit the \citetads{hoog} sample and check for the validity of the predicted birth clusters of the stars. The more so, because radial velocity studies have shown several stars of that sample to be radial velocity variable binaries \citepads{2010ApJ...721..469J}. 

\subsection{Radial velocities}

Since radial velocities are one of the 6 parameters required for the kinematic analysis, they have to be updated as well. \textit{Gaia} does not provide radial velocities for hot stars. Therefore, ground based spectroscopic studies are necessary. In order to identify close binaries showing radial velocity variations multi-epoch spectroscopy is required to pin down the orbit and in particular the systemic velocity.
\citetads{2010ApJ...721..469J} found five stars, previously suggested to originate from the Sco-Cen OB association, to be radial velocity variable and, hence, to be binaries. Therefore, we have to replace the radial velocities adopted by \citetads{hoog} by their system velocities (see Table \ref{tab:binarity}). The space velocity of the stars  (with respect to Galactic rotation) as calculated from the listed radial velocities, \textit{Gaia} EDR3 astrometry, the circular velocity of the disk at solar distance and the LSR velocity of Model I of \citetads{andreassmass} are also given in Table \ref{tab:binarity}. 

Concerning radial velocities of relevant stellar clusters data are unfortunately scarce (see Table \ref{tab:clusterparams}), for about a dozen cases they are based on one to four cluster members, only. In those cases the uncertainties may be larger than anticipated. 

\begin{table}
\begin{center}
\caption{Radial velocities v$_\textrm{rad}$ and derived space velocities v$_\textrm{s}$ for the sample of \citetads{hoog}.  }\label{tab:spacevelocity} 
\renewcommand{\arraystretch}{1.3}
\begin{tabular}{c c c c c } 
\hline
\hline
HIP& RVV & v$_\textrm{rad}$  & v$_\textrm{s}$&Reference\\
   &     &  km\,s$^{-1}$ &   km\,s$^{-1}$  &  \\
\hline
\multicolumn{5}{c}{Walkaway stars} \\
26241&Y&$28.7\pm1.1$&$5.79^{+1.01}_{-0.96}$\tablefootmark{b}&2\\
46950&Y&$16.5\pm3.0$&$9.28^{+2.12}_{-1.52}$&5\\
69491&Y&$-10.3\pm6.9$&$9.30^{+5.57}_{-2.85}$&5,7\\
42038&Y&$22.0\pm3.0$&$10.52^{+2.56}_{-2.27}$&5\\
38455&Y&$29.5\pm7.4$&$13.00^{+7.02}_{-6.09}$&6\\
48943&N&$19.4\pm0.8$&$18.58^{+0.98}_{-0.92}$&5\\
3881&Y&$-23.9\pm0.9$&$22.16^{+0.87}_{-0.87}$&2\\
38518&N&$41.0\pm4.1$&$23.31^{+4.06}_{-4.05}$&2\\
76013&Y&$16.5\pm5.0$&$24.79^{+3.85}_{-3.36}$&5\\
49934&N&$31.0\pm3.0$&$25.37^{+2.88}_{-2.66}$&2\\
57669&Y&$33.0\pm7.9$&$29.31^{+7.75}_{-7.67}$\tablefootmark{a}&3, 4\\
82868&N&$19.7\pm9,5$&$29.51^{+9.01}_{-8.61}$&4\\
\hline
\multicolumn{5}{c}{Runaway stars}  \\
81377&N&$12.2\pm3.3$&$37.29^{+2.38}_{-2.23}$\tablefootmark{b}&3\\
14514&N&$21.7\pm1.5$&$52.24^{+0.85}_{-0.79}$&2\\
39429&N&$-23.9\pm2.9$&$55.13^{+2.46}_{-2.39}$\tablefootmark{b}&2\\
102274&N&$-3.8\pm3.5$&$56.69^{+1.22}_{-1.00}$&2\\
29678&N&$36.9\pm0.7$&$62.17^{+1.60}_{-1.48}$&2\\
18614&Y&$65.4\pm0.8$&$64.54^{+9.59}_{-2.35}$\tablefootmark{a}&2\\
109556&N&$-75.1\pm0.9$&$72.30^{+1.15}_{-1.05}$&2\\
22061&N&$9.0\pm4.4$&$96.66^{+1.73}_{-1.66}$&2\\
24575&N&$56.7\pm0.6$&$107.38^{+1.0}_{-0.97}$&2\\
27204&N&$109.0\pm1.8$&$113.97^{+2.79}_{-2.57}$&2\\
91599&N&$29.2\pm0.2$&$122.98^{+2.60}_{-2.47}$
&2\\
82286&N&$149.6\pm0.6$&$161.94^{+11.65}_{-2.01}$&1\\
\hline
\end{tabular}
\tablefoot{The median of the space velocities with $1\sigma$ uncertainties are derived using \textit{Gaia} EDR3 astrometric data  \citepads{2021A&A...649A...2L} unless noted otherwise. For radial velocity variable binaries (RVV=Y) the system velocity is given.
\tablefoottext{a}{Using \textit{Gaia}   DR2 astrometry \citepads{2018A&A...616A...2L}}, 
\tablefoottext{b}{Using the new \textit{Hipparcos} reduction \citepads{2007ASSL..350.....V}}}
\tablebib{ (1) \citetads{Markus:Thesis:2021}; (2) \citetads{hoog}; (3) \citetads{2018AN....339...46Z};(4) \citetads{2020MNRAS.498..899N}; (5) \citetads{2010ApJ...721..469J};(6) \citetads{2011MNRAS.410..190T};(7) \citeads{2008MNRAS.385..381B}}


\label{tab:binarity}
\end{center}
\end{table}

  Furthermore, lower space velocities resulted from the new radial velocities and EDR3 measurements for many stars. 
  

The binaries HIP 3881, HIP 38455, HIP 42038, HIP 46950, and HIP 76013 are likely walkaways too, rather than runaway stars, because their space velocities do not exceed 30 km\,s$^{-1}$. For the same reason three single stars (HIP 38518, HIP 48943, and HIP 49934) should be considered to be walkaways as well. The new space velocity of HIP 69491 even decreased to 9.30 km\,s$^{-1}$, meaning that the star may just be a very slow walkaway star.  
Hence, eleven stars of the sample are confirmed as runaways only, of which three have space velocities exceeding 100 km\,s$^{-1}$. A more appropriate criterion is the relative velocity with respect to the parent cluster, which can be calculated only when the cluster has been identified.
They are listed in Table \ref{tab:identification}. Accordingly, the classification as walkaway or runaway star remains unchanged.


\subsection{The cluster origins}



We ran our p-value based distance minimization procedure for the runaway stars and the parent clusters suggested by \citetads{hoog}. In appendix C (Table.\ref{tab:table1})
we demonstrate that we can reproduce the results of \citetads{hoog}, if we use the same input data. These results do not hold up when old radial velocities and new \textit{Gaia} values are used for previously known single and binary stars as shown in Table \ref{tab:oldtars}. Similarly, we show that making use of the \textit{Gaia} data and more recent radial velocity measurements changes the conclusions in many cases considerably. 


\subsection{Identification of parent clusters and associations}\label{origin}
   

Using our new method and equipped with accurate astrometry along with new radial velocities we searched for candidate clusters.
We have used the best values available from the \textit{Gaia} DR2, EDR3 \citepads{2021A&A...649A...2L}, and the new reduction of \textit{Hipparcos} \citepads{2007ASSL..350.....V} and searched through more than 1000 clusters compiled from \citetads{2019ApJ...870...32K},\citetads{GaiaHR},\citetads{OCDR2},\citetads{RAVEOC},\citetads{DR2OC}, and \citetads{DR2OC2}.

Furthermore, important indicators of stellar ejection from the cluster and the ejection mechanism are the stellar age, the kinematic age, and the cluster age. We compiled age estimates of multiple clusters. The results are summarized in Table \ref{tab:identification} and shall be discussed for each individual star in the following. To estimate errors on the time of flight from these clusters we use the approximation:
\begin{equation}
    T_\textrm{rel}=\frac{D_\textrm{rel}}{V_\textrm{rel}}
\end{equation}
where D$_\textrm{rel}$ and V$_\textrm{rel}$ are the relative distance and the relative velocity between the star and the cluster. We sampled the radial velocities, proper motions, and parallaxes 10000 times for ever system, converted the coordinates to a Galactocentric Cartesian system, and looked at the star and cluster system within the frame of reference of the cluster. This allowed us to derive an estimate of the standard deviation of the time of flight of the system, assuming that a straight line connects the two objects in this frame of reference.

Finally, while our simulations take into account Gaia correlations between the parallax, the proper motion component in right ascension, and that in declination
from \citepads{2018A&A...616A...2L,2021A&A...649A...2L}, the fitting procedure is not equipped for this yet, due to the fact that we do not sample the parameters but rather fit them. Using mock data we demonstrate in appendix \ref{sect:appendix:B} that the impact of the correlations on the simulation results is negligible 
at the \textit{Gaia} precision level. Nevertheless, for objects where correlations are more than $\pm$0.3 between any two parameters we also provide simulations of trajectories to proof that the trajectories for these stars are not affected. Furthermore, none of our stars (for which a corresponding cluster was found) had a correlation factor between any two parameters beyond $\pm$0.5. 


\section{The Hoogerwerf et al. sample revisited: Individual objects}\label{sect:runaway}

We shall discuss the individual objects and their suggested parent clusters grouped into Walkaway stars (Sect. \ref{sect:walkaway}) and runaway stars (Sect. \ref{sect:runaway})

 \subsection{Walkaway stars and their suggested parent clusters}\label{sect:walkaway}
 
 In the following we shall discuss the walkaway stars and their suggested parent parent clusters individually, arranged according to in increasing space velocity (see Table \ref{tab:spacevelocity}). 

\subsubsection{HIP 26241 ($\iota$ Ori)}

HIP 26241 ($\iota$ Ori) is an excentric binary moving with a velocity of about 10 km\,s$^{-1}$ at almost right angles with respect to AE Aur/$\mu$ Col. The very different masses of the components hint at different ages of the components. N-body simulations by \citetads{2004MNRAS.350..615G} suggested that an exchange interaction occurred in an encounter between two low eccentricity binaries with comparable binding energy, which caused an older star to be swapped into the original $\iota$ Ori binary and the system to be ejected from the ONC. We are unable to confirm the place of origin (the ONC) as well as any encounter with the star AE Aurgiae. This difference occurs due to the change of parallax of $\iota$ Ori from 2.46 mas in the Hoogerwerf sample to 1.4 mas  in the second reduction \citepads{2007ASSL..350.....V}. \footnote{The case for which $\iota$ Ori was close to AE Aur occurs for a common origin 2.94 Myrs ago for the two stars with a high p-value of 0.73 only when Hipparcos data is used for both stars. In the \textit{Gaia} catalog}, AE Aur shifts from 1.74 mas to 2.57 mas in parallax, and the encounter no longer holds.

\subsubsection{HIP 46950}
HIP 46950 is a binary star, and therefore the original BSS scenario might not hold for it 
as pointed out by \citetads{2010ApJ...721..469J}. We find that it may have originated in the Gama Vel/ Pozzo~1 cluster in the Vel OB2 association as a walkaway star, just like the walkaway star HIP 38518. This cluster is 13 Myrs old, which is about the same time we find for the kinematic age of this star, making the dynamical ejection scenario more probable.

\subsubsection{HIP 69491}

The eclipsing binary HIP 69491 was suggested to come from the Upper Centaurus Lupus association (UCL) or Cep OB6. However, the high space velocity of 77 km\,s$^{-1}$ was puzzling for a binary system. Therefore, neither BSS nor DES could be confirmed. The findings of \citetads{2008MNRAS.385..381B} for a radial velocity of v$_{\textrm rad}$=$-$10.3 km\,s$^{-1}$ resolved one of the problems with this star system. It is no longer a runaway candidate. As the third slowest star in our sample, 
the star barely qualifies as a walkaway. We find two major clusters as possible origin: NGC 6475 and C1439-661. The corresponding kinematic age is 16 Myrs or 11 Myrs, respectively\footnote{We stick with Gaia EDR3 here even though the distance of d=339 pc found by \citetads{2008MNRAS.385..381B} is closer to the \textit{Hipparcos} parallax than to the {\textit  Gaia} ones}. Since NGC 6475 is almost 300 Myrs old and C1439-661 is 65 Myrs old (\citeads{2019A&A...623A.108B}), only C1439-661 qualifies as a possible parent cluster. Accordingly, the most probable mechanism is BSS.

\subsubsection{HIP 42038}
The star HIP 42038 was suspected to come from either the Upper Centaurus Lupus association or IC 2391 as a BSS candidate. Due to it being a radial velocity variable, these origins had to be called in question. We show that the evidence for IC 2391 to be a cluster of origin is flimsy when the new radial velocity is used, as also found by \citeads{2010ApJ...721..469J}. Instead, we find three other clusters which may be the origin. These are Cl VDBH 99, NGC 2451A, and Cl Platais 9. The flight times are much higher than expected, being 34, 36, and 20 Myrs respectively. The clusters themselves are older, with Cl VDBH 99 being 81 Myrs, NGC 2451A being 44 Myrs old, and Platais 9 being 78 Myrs old {\citeads{2018MNRAS.473..849D}, \citeads{2019A&A...623A.108B}}. Due to the difference of 10 Myrs between the age of the cluster and the flight time, and the high p-value of $\approx$0.96, NGC 2451A is the best candidate cluster for this star. However, in all three cases, the probability of HIP 42038 being a BSS is higher due to the fact that the kinematic age is much less than the age of the cluster. Such a scenario is possible if a third star in the system underwent a supernova. The resulting binary in this case would be a walkaway. This scenario was proposed by \citeads{hoog} to explain the evolution of HIP 69491, an eclipsing binary. We conclude that it may also apply for HIP 42038. 

\subsubsection{HIP 38455}
\citetads{2011MNRAS.410..190T} claimed that the star HIP 38455 is not a runaway since it is a binary with a systemic radial velocity of 29.5 km\,s$^{-1}$ as opposed to -31 km\,s$^{-1}$ used by \citetads{hoog}. We use the new radial velocity and confirm that it does not come from Collinder 135 as previously  suggested. However, due to its low space velocity of $\approx$13 km/s, it might be a walkaway star which came close to the moving group Platais 3 about 18 Myrs ago. The cluster is relatively old but its age is poorly known \citepads[208$^{+122}_{-41}$Myrs,][]{2019A&A...623A.108B}. Despite the age difference the cluster should not be discarded before a better age estimate becomes available.

\subsubsection{HIP 48943}
HIP 48943 is a B-type emission line star with a high projected  rotational velocity of 190 km\,s$^{-1}$ \citepads{2018A&A...609A.108S}, which was thought to come from the Lower Centaurus Crux (LCC). Due to the new radial velocity, this does not hold true anymore. We find the cluster NGC 6405 intersected the trajectory of the star 25 Myrs ago with a p-value of 0.37. The cluster is 48 Myrs old \citepads{Liu}. The stellar age is not known, but the high projected rotational velocity of the star gives credence to the BSS. Its low relative velocity makes it a walkaway star.

\subsubsection{HIP 3881 ($\nu$ And)}
For the spectroscopic binary HIP 3881, we find the cluster Cl Alessi 20 as a possible site of origin. The estimated time of flight is 15 Myrs based on EDR3 values for the star\footnote{The cluster has 3 RV studies, resulting in somewhat different radial velocities. We find a minimum distance of 20 pc using the value of v$_\textrm{rad}$=-5.04 km\,s$^{-1}$ from \citepads{OCDR2}. The encounter reported here uses v$_\textrm{rad}$=-11.5 km\,s$^{-1}$ based on \citetads{RAVEOC}. We find an even more significant encounter with a p-value of 0.73 when a RV of v$_\textrm{rad}$=-14.89 km\,s$^{-1}$ from \citetads{clusterRV2019} is used. However, the latter is only based on 1 star and, therefore should be considered too uncertain to be useful}. \citetads{2021ApJ...912..165R} estimate the age of the cluster at $9\pm4$ Myrs, while \citetads{2013A&A...558A..53K} give an estimate of 38 Myrs. Therefore, it is possible that Cl Alessi 20 is the original cluster of the star. Since the star is a walkaway binary, it may have been created in a BSS. Depending on which age estimate for the cluster is selected, it could have a DES origin as well.

\subsubsection{HIP 38518}

While Gam Vel was the prime candidate for this star as was suggested by \citeads{hoog}, we find the cluster Gulliver~9 to have also crossed paths with the star almost 8 Myrs ago. This is a more significant approach than the one to Gam Vel when EDR3 is considered. However, Gulliver 9 is also part of the Vel OB2 region, meaning that the star was ejected from one of the regions of the Vel OB2 association. The original suggestion of BSS is the likely scenario in this case due to the fact that both the clusters are slightly older than the star. Its low relative velocity (see Table \ref{tab:clusterparams}) makes it a walkaway star.

\subsubsection{HIP 76013}
The star HIP 76013 is a walkaway Be type star with an He-SdO companion. According to the mass estimates for the He-SdO \citepads{2021AJ....161..248W}, the stars would be about 30 Myrs old. We find no possible candidate clusters for this star using \textit{Gaia}  data. To check whether there may be a possible candidate with \textit{Hipparcos}, we rerun the procedure and find that the star was close to the cluster M 67, 26 Myrs ago with a p-value of 0.51. The probable origin of this binary star would be the DES if the stellar age were to be close to 26 Myrs. However, the cluster is more than 3.5 billion years old \citepads{2021A&A...645A..42I} and, therefore, it can not be a possible origin. The only other cluster we find with \textit{Hipparcos} is the cluster Blanco 1, which was close 36 Myrs ago with a p-value of 0.166. The cluster itself is about 100 Myrs old, which may rule out an origin. However, due to the fact that the companion is an He-SdO, the visible star may have gone through a rejuvenation phase via mass transfer, and could be older than 30 Myrs. A scenario of rejuvenation has been proposed before, for instance by \citetads{2010ApJ...719L..23B} to explain the origin of the hyper-velocity star HE 0437-5439.

\subsubsection{HIP 49934}

For the Be star HIP 49934, an origin in IC2602/IC2391 was suggested within the context of BSS. We find this origin does not hold, and find a different cluster NGC 3532, which intersected the star's motion 5 Myrs ago. However, the age of the cluster has been determined to be 400 Myrs \citepads{2019A&A...623A.108B}, which is higher than our cut-off age. In addition, it has been found to host white dwarfs \citepads{2021A&A...645A..13P}, cooling times seem to be consistent with the isochrone age. According to its low space velocity the star is a walkaway. The BSS does seem to be more probable in this case, as supported by the high projected rotational velocity of 220 km\,s$^{-1}$ \citepads{2020MNRAS.493.2528B}
 but the true cluster of origin is yet to be found.

\subsubsection{HIP 57669}

The Be star HIP 57669 was thought to originate in the Upper Scorpius region in the cluster IC 2602. Due to the age of the cluster, the flight time, and the high projected rotational velocity, the star was thought to be a BSS candidate, although no pulsar was traced back to it. The origin in IC 2602 was shown to not hold with new \textit{Gaia}  values \citepads{supernovaexample}. We now find that its path crossed with the young moving group $\eta$ Cha in the Chameleon star forming region 4-5 Myrs ago.
HIP 57669 travels slowly away from $\eta$ Cha (20 km\,s$^{-1}$), which makes it a walkaway from that cluster. The age of the cluster is $11\pm3$ Myrs \citepads{2021AJ....161...87D,2015MNRAS.454..593B}, suggesting that the origin could indeed be a BSS origin. This would mean it was originally a triple system as \citetads{chini} found that HIP 57669 is a spectroscopic binary. 

The star might also originate from the Cl Alessi 13 moving group, which is nearby the $\eta$ Cha moving group, although the p-value is less than the $1\sigma$ limit.
The flight time would be more than 10 Myrs, compatible with the cluster age of about 40 Myrs \citepads{2019ApJ...887...87Z}. This would make HIP 57669 a BSS ejection, as originally suggested.

 \subsubsection{HIP 82868}

\begin{figure*}[htp]
\hspace{-0.5cm}

\centering
\includegraphics[width=1.0\textwidth]{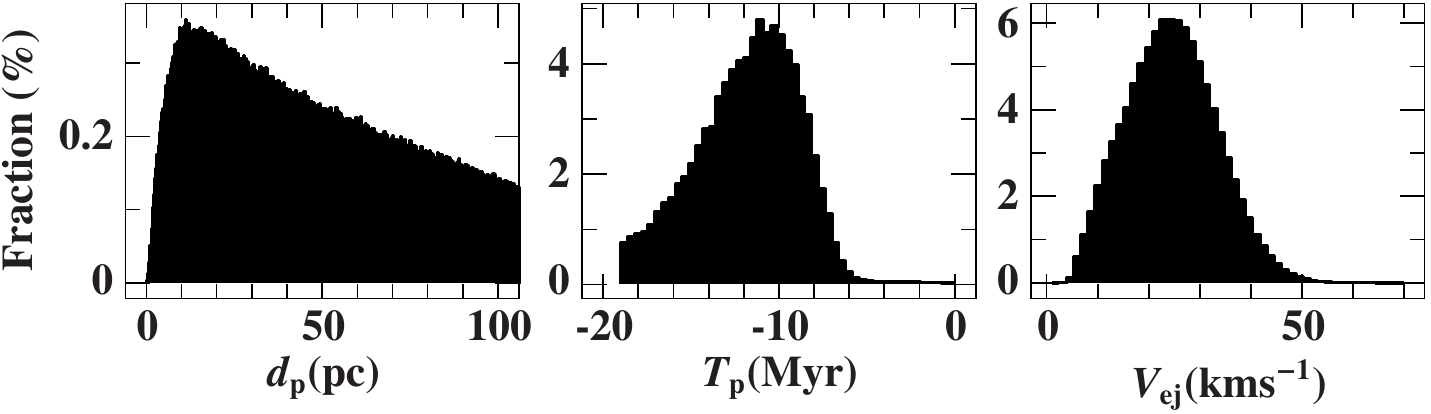}
\caption{Histograms for the Monte Carlo simulations of back-traced trajectories for HIP 82868 and Alessi 13. The left most panel shows the fraction of distances of closest encounters between the star and the cluster. The middle panel is the histogram of the times for the closest encounters of the two stars, and the right most panel plots the relative ejection velocity of the star with respect to Alessi 13. }\label{fig:histogram}
\label{fig:hip82868}
\end{figure*}

For the single Be star HIP 82868, we find the moving group Alessi 13/$X^1$ Fornacius as a place of origin. This is a well studied cluster due to its proximity to Earth ($\approx 100$ pc), and new \textit{Gaia}  data has put the cluster age at 40 Myrs \citepads{2019ApJ...887...87Z}. The flight time of 12.5 Myrs provides evidence for this star to be a BSS candidate.  This is in agreement with the original suggestion, as well as the high projected rotational velocity of 370 km/s \citepads{2010ApJ...721..469J}. The star is a walkaway traveling at 27 km\,s$^{-1}$ from Alessi 13, and due to a high correlation the simulated trajectories are provided in Fig. \ref{fig:hip82868}.
While this group may also be one of the candidates for HIP 57669, the two stars themselves never come close enough to each other with  \textit{Gaia}  values. 

\subsection{Runaway stars and their suggested parent clusters}\label{sect:runaway}

In the following we shall discuss the runaway stars in the Hoogerwerf sample and their suggested parent cluster indiviudually arranged according to in increasing space velocity (see Table \ref{tab:spacevelocity}).

\subsubsection{HIP 81377 ($\zeta$ Oph)}
 
 \citetads{hoog} suggested that the runaway star $\zeta$ Oph was ejected in a BSS event in the Upper Scorpius association and identified the pulsar PSR J1932+1059 as the runaway neutron star remnant of the former primary. A kick velocity of $\approx$350km\,s$^{-1}$
to the neutron star disrupted the binary in the supernova explosion.
 Recently, however, \citetads{supernovaexample} suggested that $\zeta$ Oph is linked to the pulsar B1706$-$16 in the association of stars known as the Scorpius-Centaurus-Lupus. This conclusion is supported by our calculations, although we find a slight mismatch of the best possible times, since a common origin with PSR J1932+1059 was found 1.3 Myrs ago with a p-value of 0.41. However, the RV of the PSR is not well constrained. This makes it possible for the time of flight to be less, and therefore more cannot be said on the subject till then.

\subsubsection{HIP 14514 (53 Ari)}

The $\beta$ Cep pulsator HIP 14514 is a classical runaway star \citepads{runawaysblaauw2} and is confirmed to come from the Orion OB1 association, which it left almost 5 Myrs ago. Due to the age of the cluster it is possible that 53 Ari was released as part of a binary system, but more data on the age of the star is required before this can be confirmed.  
 

 \subsubsection{HIP 39429}
 
We find that the O4 supergiant
 $\zeta$ Pup did indeed pass Trumpler 10, 2.5 Myrs ago, which is consistent with its young age \citepads[3.2$^{+0.4}_{-0.2}$ Myrs][]{2012A&A...544A..67B}.
 Trumpler 10
 is 55 Myrs old \citepads{2019A&A...623A.108B}, which is more than the 30 Myrs limit. However, \citetads{hoog} pointed out that $\zeta$ Pup may be a blue straggler, in which case a BSS origin in Trumpler 10 can not be ruled out.

 
 \subsubsection{HIP 102274}

The B5 star HIP 102774 was thought to originate in the Cep OB2 association. We are unable to confirm this within $2\sigma$ certainty. 
 We find two possible clusters, Alessi-Teutsch 5 and NGC 7160, from where the star might have originated. The two clusters are around 12.5 and 9.0 Myrs old (\citeads{2019AJ....158..122K},\citeads{2018MNRAS.473..849D},\citeads{RAVEOC}), meaning that the originally suggested BSS holds. 

\subsubsection{HIP 18614 ($\xi$ Per)}

HIP 18614  is a classical runaway star \citepads{runawaysblaauw2}, which is confirmed to come from the Per OB2 association via a DES ejection.

 \subsubsection{HIP 109556}
 
 We were unable to confirm the origin of the blue supergiant HIP 109556 ($\lambda$ Cep) from the Cep OB3 association when using newer data from \citetads{RAVEOC} and \citetads{2018MNRAS.473..849D}. However, using data from \citetads{2009MNRAS.400..518M} allowed us to find a close encounter 2.95 Myrs ago, which is less than that suggested by \citetads{hoog} and more in line with the expected age of the supergiant. However, Cep OB3b is less than 2.5 Myrs old \citepads{2018MNRAS.477..298G}, meaning BSS would not hold. Therefore, Cep OB3a is a more likely parent.

\subsubsection{HIP 22061 and HIP 29678}

HIP 22061 and HIP 29678 are prime candidates for a dynamical ejection, from the cluster Collinder 69. However, the stars never come closer than 7 pc, which is incidentally also the radius within which 50\% of the stars in Collinder 69 reside. Therefore, the cluster  Collinder 69 remains the best candidate for these two stars. While \citetads{hoog} claimed the two stars were not related to the supernova that formed the ring around $\lambda$ Ori, newer results have shown this explosion to be about 1 Myrs ago \citepads{2002AJ....123..387D}. This is the same as the flight times of HIP 22061 and HIP 29678. Since the stars come from the slightly outer regions of the cluster, they may have been involved with this supernova.

\begin{figure*}[htp]
\hspace{-0.5cm}

\centering
\includegraphics[width=1.0\textwidth]{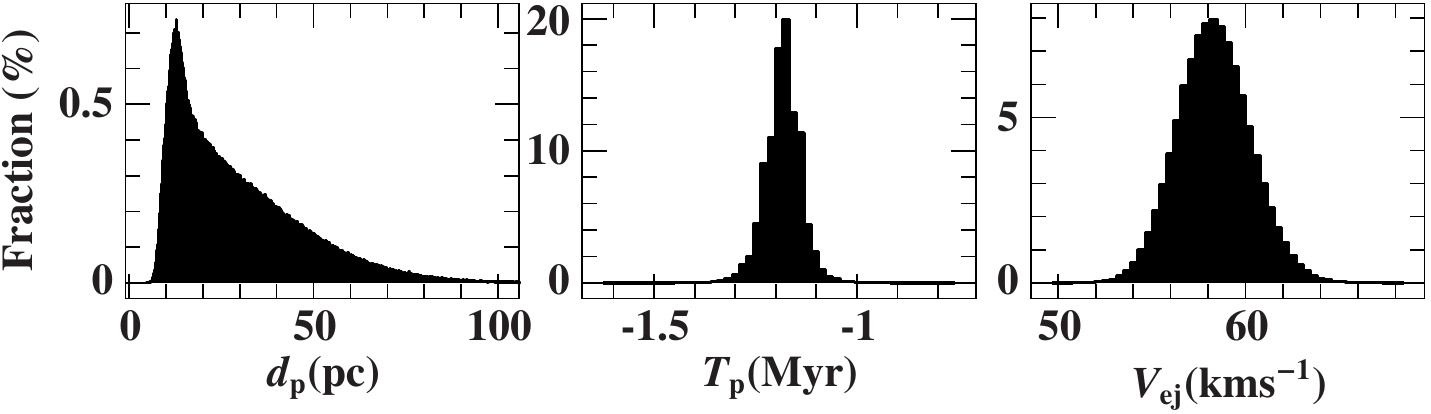}
\caption{Histograms for the Monte Carlo simulations of back-traced trajectories for HIP 29678 and Collinder 69. The left most panel shows the fraction of distances of closest encounters between the star and the cluster. The middle panel is the histogram of the times for the closest encounters of the two stars, and the right most panel plots the relative ejection velocity of the star with respect to Collinder 69. }\label{fig:histogram}
\label{fig:hip29678}
\end{figure*}

For HIP 29678, Fig. \ref{fig:hip29678} illustrates the histograms for close trajectories. Like our fitting procedure predicted, the simulations have almost no trajectories with closest distances lower than 5 pc.

 \subsubsection{HIP 24575 (AE Aur), HIP 27204 ($\mu$ Col), and Collinder 69}\label{sect:aeaur}

The star $\mu$ Col (HIP 27204) has been one of the exemplary stars for the DES along with AE Aur (see Sect. \ref{aeaur}). The two stars were found to move in almost opposite directions away from the ONC with velocities of about 100 km\,s$^{-1}$. 
However, we found evidence that $\mu$ Col might instead have its origin in Collinder 69, which is also in the Orion region, and thought to be where the two stars HIP 22061 and HIP 29678 originate from. 


\begin{figure*}[htp]
\hspace{-0.5cm}

\centering
\includegraphics[width=1.0\textwidth]{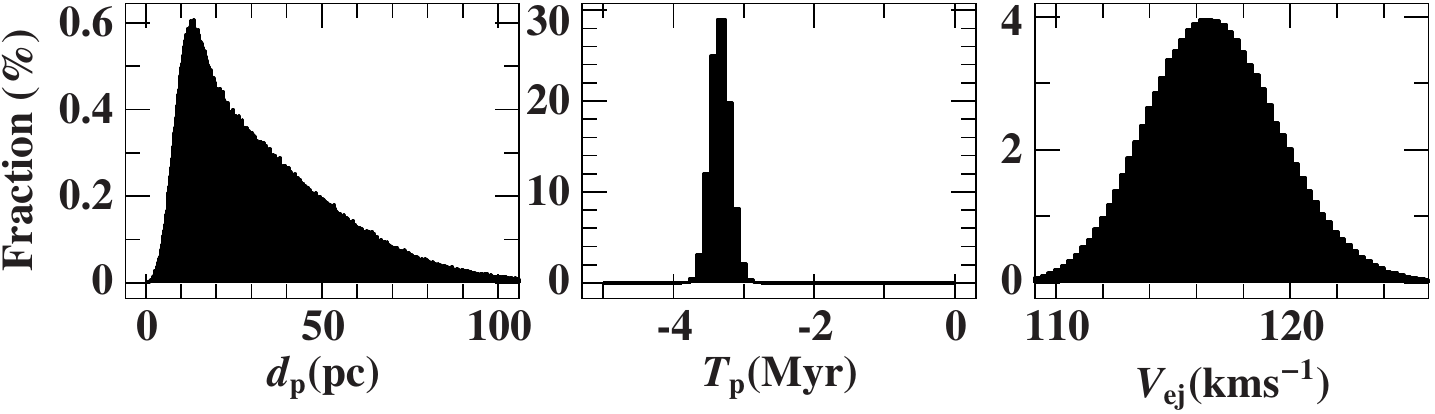}
\caption{Histograms for the Monte Carlo simulations of back-traced trajectories for $\mu$ Col and Collinder 69. The left most panel shows the fraction of distances of closest encounters between the star and the cluster. The middle panel is the histogram of the times for the closest encounters of the two stars, and the right most panel plots the relative ejection velocity of the star with respect to Collinder 69. }\label{fig:histogram}
\label{fig:mucollamori}
\end{figure*}

\begin{figure}[htp]
\centering
\includegraphics[width=0.5\textwidth]{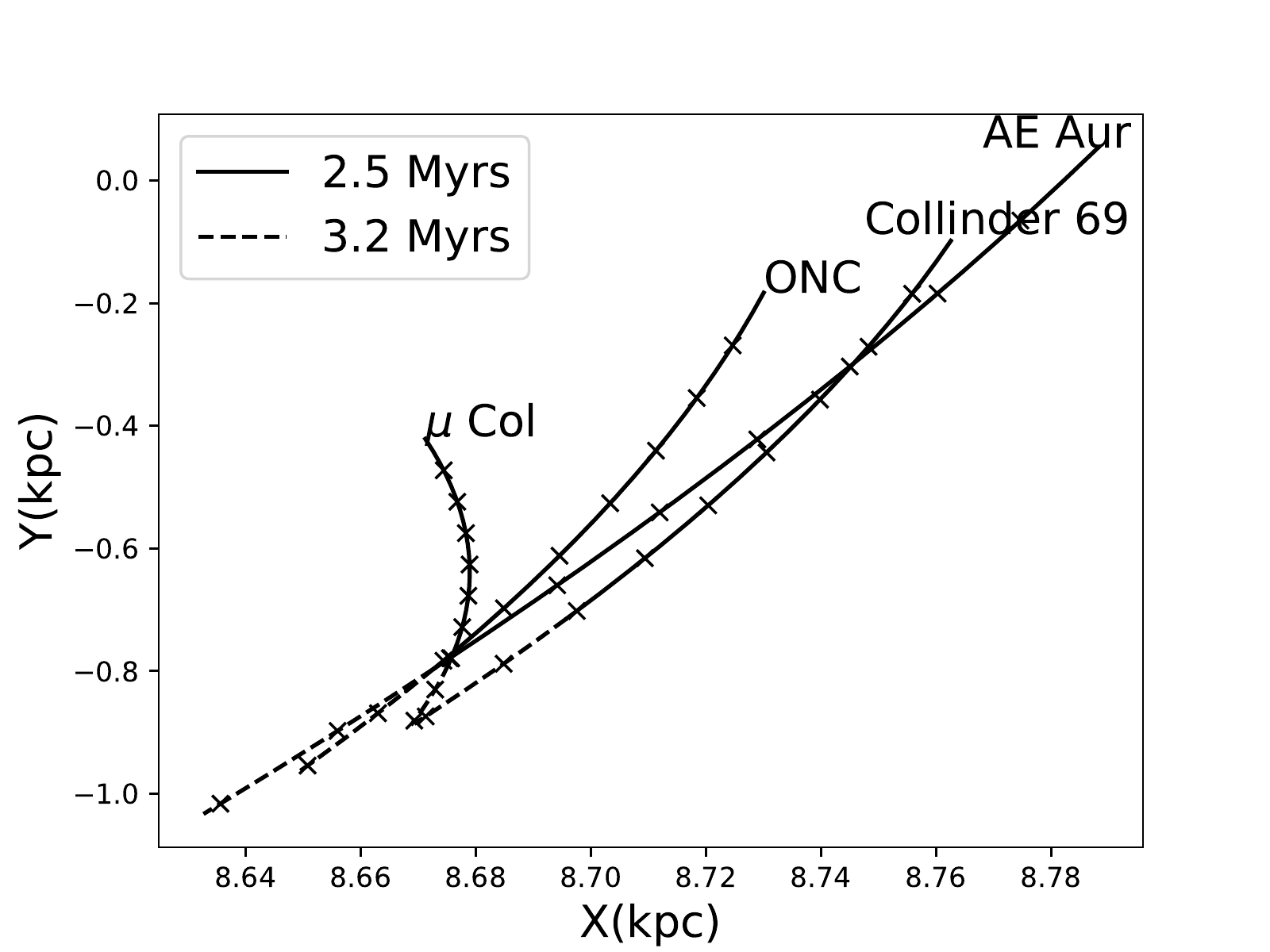}
\caption{Orbits for AE Aurigae and $\mu$ Col using best-fit parameters for their respective origins from the ONC and Collinder 69. The figure illustrates how $\mu$ Col may have been ejected from Collinder 69, and while it passed through the ONC, AE Aurigae was ejected from the latter. X and Y represent the Galactocentric Cartesian coordinates with the Sun at (8.4 kpc,0). Crosses represent 0.35 Myrs of backward trajectory. The trajectories begin at the respective labels. Trajectories are plotted using Galpy \citepads{bovy}.} 
\label{fig:orbitsaeaurmucol}
\end{figure}

$\mu$ Col has a high p-value of 0.54 for a common encounter with the Collinder 69 cluster 3 Myrs ago. The cluster is $\approx$5 Myrs old \citepads{2020ApJ...902..122K}, and a prime site for the DES mechanism as evidenced by the two other stars. That $\mu$ Col can also be traced back to the ONC is a coincidence arising from the facts that the ONC is not that far away from Collinder 69 and AE Aurigae originated in the ONC at the same time. The fact that a runaway star can be a visitor to another dense star forming region has been shown before by \citeads{2021MNRAS.501L..12S}, for the ONC itself. While they found past and future visitors to the ONC, $\mu$ Col was not considered, due to the fact that it has long been accepted in literature that it originated in the ONC\footnote{This scenario of $\mu$ Col crossing the ONC after its origin in Collinder 69 occurs only if Collinder 69 is closer to us than the ONC.}. Since the evidence for $\mu$ Col and its ONC origin have relied on Monte Carlo simulations, like the ones presented in Sect. \ref{aeaur}, we also carried out MC simulations 
for Collinder 69 and $\mu$ Col to have been at the same place. The histogram of closest encounter from a million backward trajectories for each object, along with the time of closest encounter, and the ejection velocities, are shown in Fig. \ref{fig:mucollamori}. The histograms match the spread for two objects thought to come from the same place (cf. Figs. \ref{fig:hist} and \ref{fig:mock1}), with the peak of the distribution below 20 pc.
Since this common encounter is older than 2.5 Myrs, it is more probable for $\mu$ Col to have been born in Collinder 69 instead of the ONC. The best-fit trajectories for both the stars and their respective places are illustrated in Fig. \ref{fig:orbitsaeaurmucol}.
 The average time of more than 3 Myrs provides evidence that $\mu$ Col may have originated in this cluster instead of ONC.

\subsubsection{HIP 91599}

\begin{figure}[htp]
\centering
\includegraphics[width=0.5\textwidth]{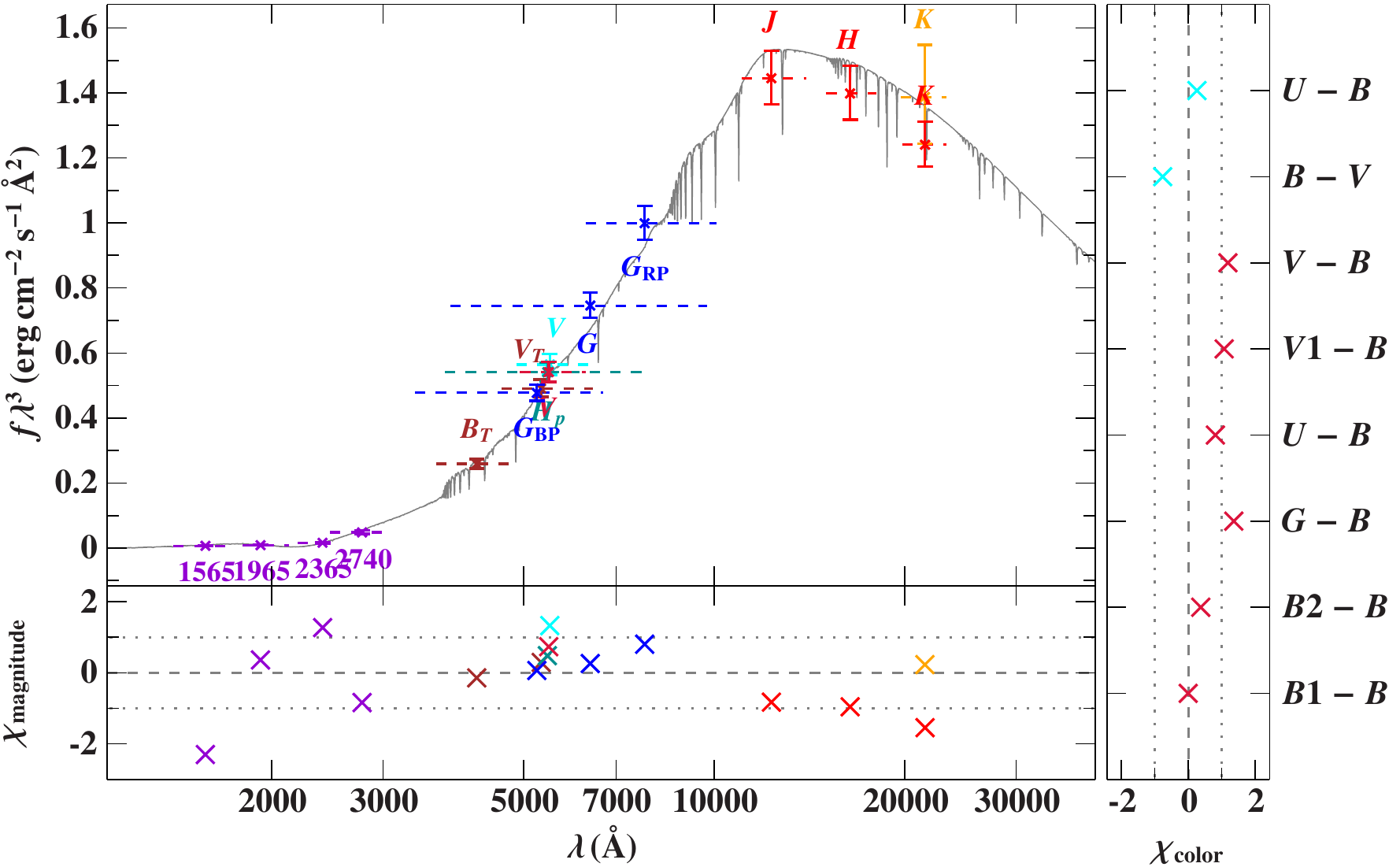}
\caption{Comparison of synthetic and observed photometry for HIP 91599. The top
panel displays the spectral energy distribution. The colored data points are filter-averaged fluxes
that were converted from observed magnitudes. The dashed horizontal lines depict the full
widths at 10\% of the maximum of the filter transmissions, while the solid gray line represents the best fitting model (spectral
resolution of 6 \AA). In order to reduce the steep slope of the SED the flux is multiplied by the wavelength to the power
of three. Bottom panel: Residuals, $\chi$: differences between synthetic and observed magnitudes divided by the corresponding uncertainties.  The photometric systems have the following color code: UV magnitudes from the TD1 satellite \citepads[violet;][]{ 1978MNRAS.184..733N} labeled with central wavelengths, optical magnitudes from Tycho2 \citepads{2000A&A...355L..27H}, \textit{Hipparcos} \citepads{2007ASSL..350.....V}, \textit{Gaia}   \citepads[blue][]{2020yCat.1350....0G}, and \citetads{2006yCat.2168....0M}; IR photometry from 2MASS \citepads[red][]{2003yCat.2246....0C} and DENIS\citepads[yellow][]{2005yCat.2263....0D}.
Right hand panel: color indices in the Johnson (cyan) and Geneva (red) systems \citepads{2006yCat.2168....0M,1988csmg.book.....R}.
}
\label{fig:sed}
\end{figure}

The star is a B-type star \citepads[T$_{\rm eff}$ =26530K, log g=3.65][]{2004ApJ...604..362D} with a projected rotational velocity of 92 km\,s$^{-1}$ 
(\citeads{2007AJ....134.1570D},
\citeads{2020yCat..36340133G}). 
Its parallax given in the \textit{Hipparcos} catalog and its new reduction \citepads{1997ESASP1200.....E} is large (3.61 $\pm $ 1.16 mas and 3.41 $\pm$ 0.72 mas, respectively). However, its parallax from both \textit{Gaia}  catalogs are much smaller (0.93 mas and 0.91 mas in \textit{Gaia}  DR2 and EDR3, respectively).
The star is in a crowded field at a very low Galactic latitude of -1.5$^\circ$.
This allows an independent distance estimate be made from reddening-distance relations. By fitting the spectral energy distribution (SED) and color indices of the star (see Fig. \ref{fig:sed}), we derived its interstellar reddening as described in \citetads{2018OAst...27...35H}. Our fit gives a color excess of E(B-V)=0.77 $\pm$0.02 mag. The reddening-distance relations of \citetads{2017A&A...606A..65C} suggest that its distance must exceed 1 kpc consistent with the \textit{Gaia}  parallax. Consequently, its space space velocity is high, making HIP 91599 the second fastest star in the disk runaway sample.

\begin{figure}[htp]
\centering
\includegraphics[width=0.5\textwidth]{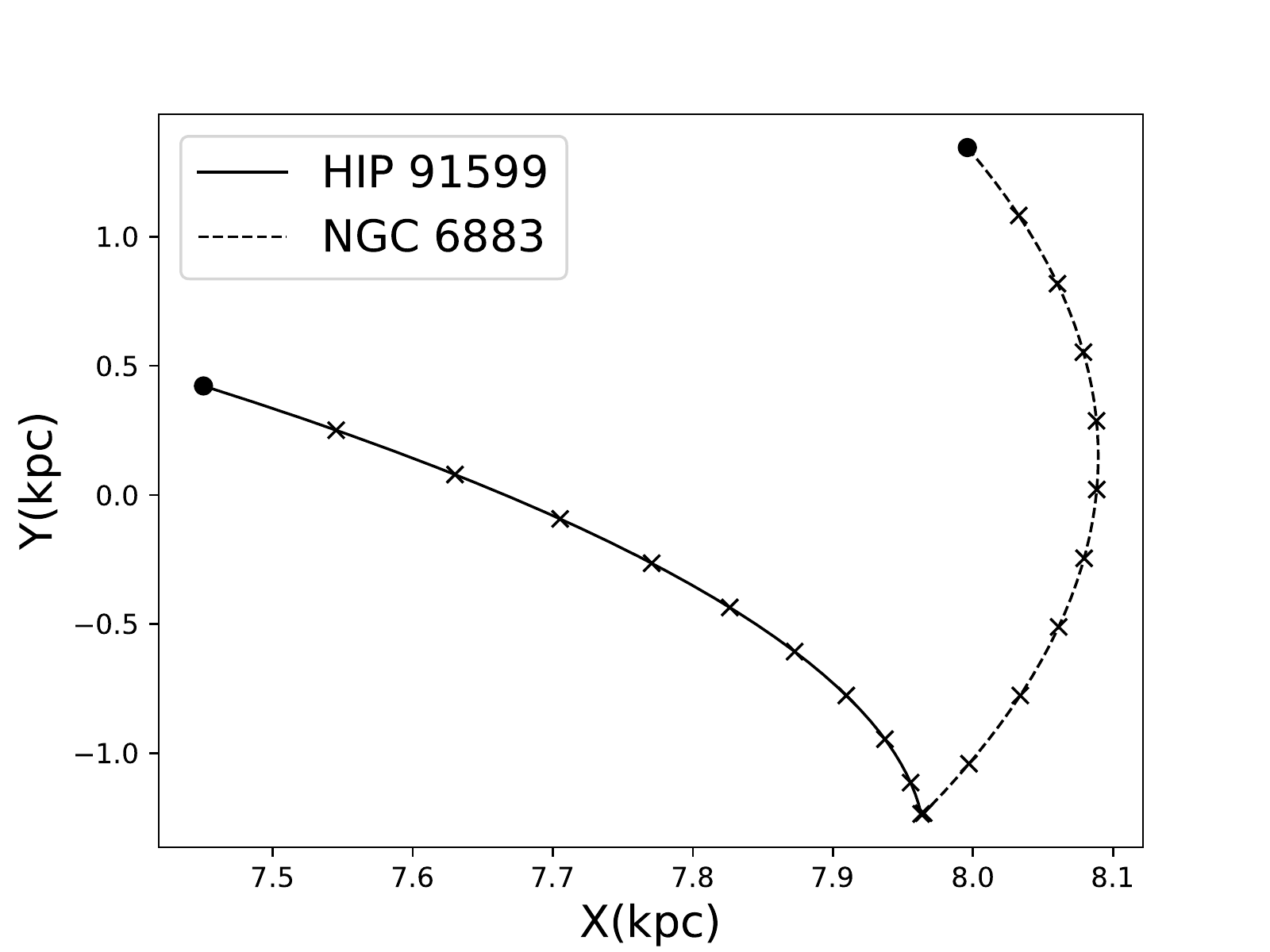}
\caption{The orbits of the star HIP 91599 and the cluster NGC 6883 for the best-fit parameters shown here for 10.7 Myrs in the past. Crosses represent 1.1 Myrs of backward trajectory and circles represent present day positions. X and Y are Galacto-centric Cartesian coordinates. Trajectories are plotted using Galpy \citepads{bovy}.
}
\label{fig:orbithip91599}
\end{figure}

HIP 91599 was thought to come from either the Per OB2 or the Per OB3 associations. 
We find that the cluster NGC 6883 has a minimized orbit with the star, 10.7 Myrs ago with a p-value of 0.92 (we adopt a distance of 1380 pc from \citetads{2018MNRAS.473..849D}, assuming a 10\% error). This cluster is around 50 Myrs old (\citeads{2018MNRAS.473..849D},\citetads{RAVEOC}), which is more than a 30 Myrs difference. However, HIP 91599 is another blue straggler candidate \citepads{hoog}, therefore making BSS a likely origin mechanism from this cluster despite the time difference. This is further supported by the high Helium abundance of the star. The best-fit orbit is shown in Fig. \ref{fig:orbithip91599} 

\subsubsection{HIP 82286 (HD 151397) - an overlooked runaway B star}

The star HIP 82286 (HD 151397) was part of the runaways already studied by \citetads{runawaysblaauw2}, who suggested the cluster NGC 6231 as a possible site of origin. The star was not included in the \citepads{hoog} sample, because \textit{Hipparcos} failed to deliver useful astrometry. The new reduction improved it, but the uncertainty of the parallax still remained at the 70\% level. Its high radial velocity of v$_\textrm{rad}$=151 $\pm 4.7$ km\,s$^{-1}$ \citepads{2012A&A...546A..61D} makes it a highly interesting object.
Hence the star was revisited by \citet{Markus:Thesis:2021} in a spectroscopic analysis of high Galactic latitude runaway stars (see Sect. \ref{sect:high_vel}), who found it to be a very young (1.3 Myrs), hot (T$_\textrm{eff}$=29760$\pm$300 K) and massive (13.8 M$_\odot$) main sequence 
star.

\begin{figure*}[htp]
\centering
\includegraphics[width=\textwidth]{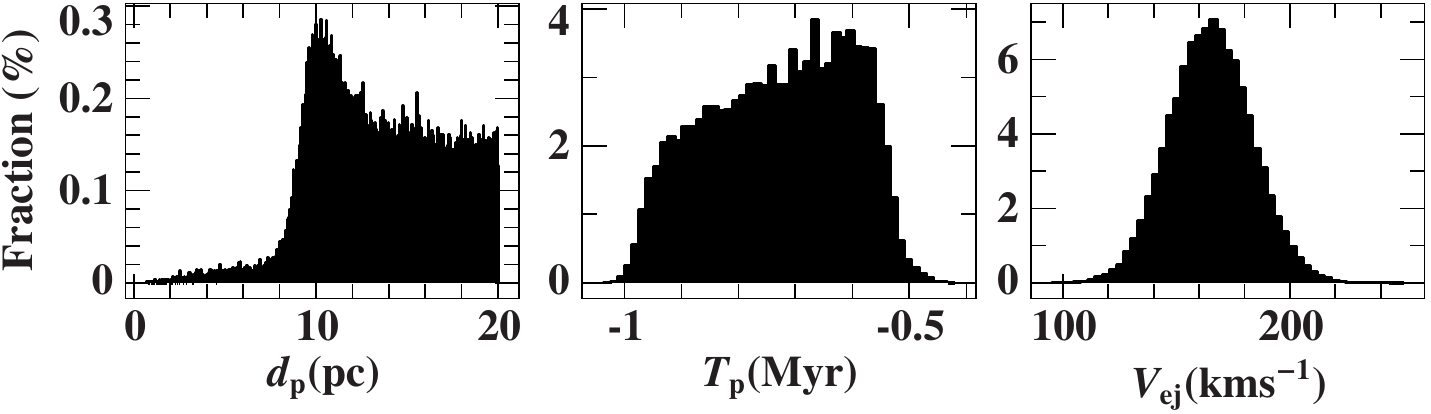}
\caption{Simulated histograms of closest encounter, time of closest encounter, and ejection velocities for HD151397 and Cl VDBH 205 for the fraction of trajectories coming closer than 20 pc.}
\label{fig:hd151397}
\end{figure*}

NGC 6231 was not confirmed as its parent cluster, and since then no possible cluster or association has been linked to the origin of this star. 
We used the radial velocity along with \textit{Gaia} astrometry to search for possible origin sites for this star.
About 0.7 Myrs ago, HD 151397 was within 7 pc of the cluster Cl VDBH 205 (in the SCO OB1 association) with a high p-value of more than 0.96 when using \textit{Gaia} EDR3 astrometry. This is slightly outside the inner part of the cluster which has 50\% stars within 3 pc. 
The kinematic age is less than the star's evolutionary age of 1.5 Myrs. The cluster is 
somewhat older. Its age is given as 13 Myrs by \citetads{2017MNRAS.470.3937S}  and 6.8 Myrs \citepads{2021A&A...647A..19T}, both, however, without uncertainties.
It is possible that the star is a blue straggler resulting from mass transfer in a binary system prior to the SN explosion. Since this star has a correlation higher than 0.3 between the proper motions, we show the simulations for close encounters in Fig. \ref{fig:hd151397}.
Therefore, we suggest that this star may have originated in Cl VDBH 205 in the BSS. 
 
%
%
%
%
%

\subsection{The \citetads{hoog} sample in the Gaia era}

The results described above for the stars of the \citetads{hoog} individually are summarized here.
Firstly, the space velocities of several stars were found to be lower than previously derived, leaving us with a sample of
12 walkaway with space velocities between 5 and 30 km\,s$^{-1}$ and 11 runaway stars with space velocities between 37 and 120 km\,s$^{-1}$. Furthermore, there are eight spectroscopic binaries among the walkaways and only one ($\xi$ Per) among the runaways. 
Parent clusters or associations have been identified for all but 4 runaway stars (HIP 76013, HIP 109556, HIP 38455, and HIP 49934). For HIP 38455 and HIP 49934 we found spatial coincidence in the past with two clusters, but the cluster ages ruled out a common origin. For HIP 109556, the only close encounter with Cep OB3 was found when using \textit{Hipparcos} second reduction based data from \citetads{2009MNRAS.400..518M}. We confirm the originally suggested parent clusters for 53 Ari, HIP 22061 and HIP 29678, AE Aurigae, $\psi$ Per, and HIP 38518.

We found that the $\iota$ Ori was not involved in a close encounter with AE Aurigae or the ONC. Furthermore, $\mu$ Col was found to have a close encounter with Collinder 69, 3 Myrs ago, earlier than its flight time from the ONC. Therefore, the classical example of dynamical ejection scenario is no longer found to hold true. 

New parent clusters were identified for 11 stars. Out of these, 9 clusters had p-values higher than the $1\sigma$ limit. The other 2 had p-values between  $1.2-1.5\sigma$ levels.
Finally, $\zeta$ Oph was confirmed to be associated with the runaway neutron star PSR B1706-16 as has recently been suggested by 
\citetads{supernovaexample}.

\begin{table*}
\begin{center}
\tiny
\caption{Suggested Clusters as places of origin for runaway stars based on spatial coincidence.
 }
\renewcommand{\arraystretch}{1.0}
\begin{tabular}{ c c c c c c c c c c c } 
\hline
\hline
HIP Num & Origin & p-val & T$_\textrm{f}$&T$_\textrm{rel}$ &V$_\textrm{rel}$& T$_\textrm{c}$& Flag$_\textrm{age}$& Suggested &Original & References\\
       &    &       & Myrs   & Myrs       & km\,s$^{-1}$ &   Myrs &          &    Mechanism            &  Mechanism & \\
\hline
3881& \textbf{Cl Alessi 20}&0.23&-15.19&$15.68\pm 4.12$&17.96&$9^{+4}_{-4}$/38&Y&DES/BSS&DES&1,2,16,17 \\

14514&Orion OB1a&0.77&-4.69&$4.96\pm 0.72$&53.06&$11.4^{+1.9}_{-1.9}$&Y&BSS/DES&BSS/DES&3,18\\

18614&Per OB2&0.69&-6.40&$9.29\pm 4.09$&47.9&$7^{+1.0}_{-1.0}$&Y&DES&BSS& 3,19\\

22061&Collinder69 &0.99\tablefootmark{a}&-1.016&$1.08\pm 0.11$&102.6&$4.7^{+6.3}_{-2.4}$&Y&DES&DES& 4,5,20\\
24575&ONC&0.96\tablefootmark{b}&-2.5&$2.58\pm 0.05$&105.8&$2^{+2.0}_{-2.0}$&Y&DES&DES&6,1,21\\
26241   & ONC & None  &  - & -  & - & &-& DES & \\
27204&ONC&0.94&-2.42&$2.47\pm 0.17$&113.93&&Y&DES&DES&\\
&\textbf{Collinder 69}&0.65&-3.2&$3.35\pm 0.13$&117.0&$4.7^{+6.3}_{-2.4}$&Y&DES&&1,2,22\\
29678&Collinder 69&0.98\tablefootmark{a}&-1.19&$1.32\pm0.15$&59.1&$4.7^{+6.3}_{-2.4}$&Y&DES&DES& 4,5,20\\
38455&\textbf{Cl Platais~3}&0.306&-17.8&$19.37\pm 7.87$&20.76&$208^{+122}_{-41}$&N&BSS&BSS&1,7,22\\
38518&\textbf{Cl Gulliver 9}&0.76&-7.95&$9.17\pm 3.61$&19.26&$14.2^{+900}_{-0.9}$&Y&BSS&BSS&1,7,22\\
&Gam Vel Subcluster &0.015&-4.92&$13.00\pm 3.54$&27.10&$13.0^{+1.9}_{-0.9}$&Y&BSS&& 8,22\\
39429&Trumpler 10&0.84&-2.6&$2.07\pm 0.37$&56.20&$55.0^{+1.4}_{-2.3}$&Y&BSS&-&1,7,22\\
42038& \textbf{NGC 2451A}&0.996&-36.561&$27.96\pm 5.82$&8.37&$44.3^{+1.6}_{-1.8}$&Y&BSS&BSS&1,10,22\\
&&0.42&-35.8&$22.73\pm 6.17$&7.68&&& & & 1,7\\
& \textbf{Cl VDBH 99}&0.64&-34.15&$18.80\pm 3.69$&8.15&$80.9^{+6.7}_{-4.7}$&Y&BSS&-&1,7,22\\
 &\textbf{Cl Platais 9}&0.30&-20.2&$32.58\pm 23.72$&9.32&$78.3^{+10.1}_{-4.0}$&Y&BSS&-&1,7,22\\
& IC 2391&0.0004&-18.78&$29.93\pm 11.37$&11.2&$36.0^{+2.0}_{-2.0}$&Y&BSS&BSS&1,7,23\\
46950& \textbf{Gam Vel subcluster}&0.73&-14.95&$13.77\pm 4.10$&5.99&$13.0^{+1.9}_{-0.9}$&Y&DES&BSS&1, 7,22\\

48943&\textbf{NGC 6405}&0.40&-24.8&$20.97\pm 2.70$&22.96&$48.0^{+2.9}_{-2.9}$&Y&BSS&BSS&1,7,23\\
49934&\textbf{NGC3532}&0.47&-4.27&$3.76\pm 0.81$&27.7&$399^{+5.5}_{-3.6}$&N&BSS&BSS&9,22\\

57669&\textbf{$\eta$ Cha Association}&0.54&-5.46&$4.93\pm 2.52$&20.7&$11^{+3.0}_{-3.0}$&Y&BSS&BSS&1,10,24 \\
&\textbf{Cl Alessi 13}&0.10&-12.61&$10.25\pm 4.88$&16.32&40&Y&BSS&&1,10,25\\

69491&\textbf{NGC 6475}&0.20&-16&$13.10\pm 1.62$&12.3 &$300^{+0.7}_{-0.}$&N&BSS&BSS&1,7,22\\
&\textbf{C 1439-661}&0.183&-11.44&$17.25\pm 5.64$&17.95&$64.5^{+13.41}_{-24.29}$&Y&BSS&&1,7,22\\

 76013&Cl Platais 10  &None&-&-&&-&-&-&BSS&1,7\\

81377&Sco-Cen OB 2\tablefootmark{c}&0.81&-0.742&$2.65\pm 0.21$&33.15&5-6&Y&BSS&BSS&12\tablefootmark{d},13,19\\
82286&Cl VDBH 205&0.96&-0.70&$1.00\pm0.71$&169&13&Y&BSS&-&1,7\\
82868&\textbf{Cl Alessi 13}&0.65&-12.35&$14.18\pm 5.86$&27.74&40&Y&BSS&BSS&1,10,25\\

 91599&\textbf{NGC 6883}& 0.92 &-10.73&$9.46\pm 1.50$&98.75&$51.2^{+7.4}_{-7.4}$&Y&BSS&BSS/DES&10,14,26\\
102274&\textbf{Alessi-Teutsch 5}&0.62&-2.7&$2.72\pm 0.29$&61.9&$12.5^{+1.6}_{-1.6}$&Y&BSS&BSS&1,10\tablefootmark{f},27\\
&&0.97&-3.03&$3.01\pm 0.23$&53.09&&&&1,7\tablefootmark{g}\\
&\textbf{NGC 7160}&0.86&-2.77&$3.11\pm 0.54$&48.58&$9.0^{+0.5}_{-0.5}$&Y&BSS&&1,10,23\\

109556&Cep OB3&$<2\sigma$&-3.70&-& 52.55&5-7&Y&-&BSS&10\tablefootmark{h},19\\
&Cep OB3& 0.11&-2.95&$2.69\pm 0.87$&64.37&&Y&BSS&BSS&12\tablefootmark{i}\\
\hline
\end{tabular}
\tablefoot{The closest approach to the cluster center is chosen to be less than 1 pc if not stated otherwise.
Identifications that differ from those of \citetads{hoog} are highlighted in boldface. Listed are the statistical p-values for best trajectories, times of flight to the parent cluster T$_\textrm{f}$ using gravitational potential, relative ejection velocities V$_\textrm{rel}$ with respect to the parent cluster, the corresponding cluster's age T$_\textrm{c}$, the time of flight T$_\textrm{rel}$ calculated assuming a straight trajectory with $1\sigma$ errors, flag Flag$_\textrm{age}$ based on whether the cluster age is in line with the suggested mechanism with 'Y' meaning it is and 'N' meaning it is not a suitable site of origin, suggested ejection mechanism by us, and suggested mechanism by \citetads{hoog}. References are given for the relevant input parameters, occasionally alternative entries are given when published values disagree. 
\tablefoottext{a}{closest approach to cluster center at 7 pc},\tablefoottext{b}{closest approach at 2.4 pc},\tablefoottext{c}{closest approach at 12 pc},\tablefoottext{d}{new \textit{HIPPARCOS} reduction  \citetads{2007ASSL..350.....V}},\tablefoottext{e}{closest approach at 18 pc},\tablefoottext{f}{radial velocity v$_\textrm{rad}$=-42.0 km\,s$^{-1}$},\tablefoottext{g}{radial velocity v$_\textrm{rad}$=-21.0 km\,s$^{-1}$},\tablefoottext{h}{distance D=0.66 kpc; PM=-1.759,-1.445 mas/yr; radial velocity v$_\textrm{rad}$=-22.9 km\,s$^{-1}$},\tablefoottext{i}{d=0.7Kpc; PM=-0.581,-3.434 mas/yr; radial velocity v$_\textrm{rad}$=-22.9km\,s$^{-1}$}
}
\tablebib{ Kinematic of clusters/associations: (1) PM,Par\citepads{DR2OC2}; (2) RV \citepads{clusterRV2019}; (3) \citetads{2017MNRAS.472.3887M} (10\% er parallax); (4) \citetads{clusterRV2019}; (5) \citetads{DR2OC}; (6)  \citetads{2019ApJ...870...32K}; (7)  RV \citepads{OCDR2}; (8) \citetads{2018MNRAS.481L..11B};(9) \citetads{GaiaHR}; (10) \citetads{RAVEOC}; (11) PSR PM from \citetads{1997MNRAS.286...81F};(12)  \citetads{2009MNRAS.400..518M}; (13) \citeads{2018MNRAS.476..381W}; (14)PM from \citetads{2014A&A...564A..79D}; (15) \citetads{2005A&A...438.1163K}; Age estimates of Clusters: (16) \citetads{2021ApJ...912..165R}; (17) \citetads{2016A&A...585A.101K}; (18) \citeads{1994A&A...289..101B};(19) \citetads{1997MNRAS.285..479B}; (20) \citetads{2021ApJ...917...21S}; (21) \citetads{2011A&A...534A..83R}; (22) \citetads{2019A&A...623A.108B}; (23) \citeads{Liu}; (24) \citetads{2015MNRAS.454..593B}; (25) \citetads{2019ApJ...887...87Z};(26) \citetads{2016A&A...593A.116J}; (27) \citetads{2019MNRAS.486.572six-dimensional}  
}

 \label{tab:identification}
\end{center}
\end{table*}

\section{Runaway B stars in the Galactic halo}\label{sect:high_vel}

The list of stars we have so far discussed are all stars with space velocities between 10 and 100 km\,s$^{-1}$, with a few slightly above 100 km\,s$^{-1}$. These stars are young and possibly still within the disk of the Galaxy. Many are  Be stars, binaries, and supergiants.
However, runaway stars far from the Galactic plane are also known \citepads{1956PASP...68..242G,1974ApJS...28..157G,1987fbs..conf..149T}. \citetads{2011MNRAS.411.2596S} studied the sample known at the time. In addition several runaway B-type stars have been discovered and analyzed since then
\citepads[e.g.][]{2017ApJ...842...32M,2019A&A...628L...5I,2018AJ....156...87L}.
Recently, a subset of main sequence B stars from that sample has been analyzed by
\citetads{Markus:Thesis:2021} using high resolution spectra and Gaia based kinematics. All stars are B main sequence stars and none is known to be a binary. We refer to his results as listed in Table \ref{tab:hyperrun} in the appendix or to \citeads{Markus:Thesis:2021}.
Due to their high velocity, these stars are farther away than the runaways discussed so far, making it difficult to trace back their orbits. One of the most important use of our fitting procedure is to find possible sites of origins of the halo runaway stars, which is presented in this section. The kinematic parameters of the stars are summarized in Table \ref{tab:hyperrun} in the appendix. Due to the fact that these stars are much farther away from us ($>1$ Kpc), we use Gaia EDR3 proper motions, but stick to spectroscopic distances from 
\citetads{Markus:Thesis:2021}.


\subsection{PG 1332+137 and Sai 132}
The B5V star PG 1332+137, also known as Feige 84, is a 5.5 M$_\odot$ star with an evolutionary age of 46 Myrs and a Galactic rest frame velocity of 299 km\,s$^{-1}$ \citepads{Markus:Thesis:2021}. 
We find only one cluster close to it. The star rotates moderately fast \citepads[140$\pm15$ km\,s$^{-1}$][]{2004MNRAS.349..821L}.
The cluster Sai 132 was close to the star with a p-value of 0.99, 20.7 Myrs ago. However, the cluster age is poorly known. According to  \citetads{2013A&A...558A..53K} the cluster is old ($\approx$500 Myrs), while \citetads{Liu} give an age of just 6$\pm$0.4 Myrs. Hence, any conclusion has to await the resolution of the cluster age discrepancy.

\subsection{PHL 159 and Cl Basel 2}

The star PHL 159 is an 8 M$_\odot$ star with an evolutionary age of 21 Myrs. 
We find the cluster Basel 2 was nearby the star with high p-value of 0.73. The time of flight of 24.9 Myrs is slightly inconsistent with the evolutionary age, a discrepancy which has been pointed out before by \citeads{Markus:Thesis:2021}. This tension may be accounted for if PHL 159 underwent a BSS ejection, which would make the star appear younger due to binary evolution. The cluster Basel 2 however is much older than the star \citepads{2010AstL...36...14G}, making it inappropriate for such a mechanism.

\subsection{PB 5418 and Cl Gulliver 43}
PB 5418 is a 6.5 M$_\odot$ star with an evolutionary age of 42 Myrs old. It has a radial velocity of v$_\textrm{rad}$=146.6 $\pm 2.6$ km\,s$^{-1}$. We find that it crossed paths with the open cluster Gulliver 43 20 Myrs ago with a p-value of 0.67. The age of the cluster is not known but the time of flight and evolutionary age suggest that BSS is the more probable scenario in this case. 

\subsection{HIP 70275 and NGC 1502}

The 9 M$_\odot$ star HIP 70275 is $\approx$17 Myrs old and has a radial velocity of v$_\textrm{rad}$=242.3 $\pm 0.2$ km\,s$^{-1}$. The star comes close to the young cluster NGC 1502 with a p-value of 0.52, 11.2 Myrs ago. The age of the cluster is also around $11\pm 0.2$ Myrs \citepads{2018AJ....155...91Y}. Along with the low projected rotational velocity 
\citepads[v$_\textrm{rot}$=64.3$\pm$3 km\,s$^{-1}$][]{2017ApJ...842...32M} 
a DES ejection would be most probable if the cluster is indeed the correct one. Otherwise the difference in evolutionary age and flight time suggests a BSS ejection, which would not have been possible in NGC 1502, again assuming the cluster age is correct. \footnote{There are many other clusters which are also in this region and were close to the star in the past like NGC 1513, Cl Berkeley 9, NGC 1664, and Cl Berkeley 67. However, all of them are older than 350 Myrs old according to \citetads{Liu}}.  




\subsection{HIP 11809 and [FSR2007] 0883}
 HIP 11809 (Feige 23) was close to the cluster [FSR2007] 0883, 10.2 Myrs ago respectively. 
 The estimated stellar age of 100 Myrs, and the high projected rotational velocity of 240 km\,s$^{-1}$ \citepads{2004AJ....128.2474M} implies a BSS origin. The cluster is older and the published ages range from 234 Myrs
 \citepads{2014A&A...564A..79D}
 to 700 Myrs \citepads{Liu}. Therefore it may not be the origin cluster of the star, if the cluster age estimates are correct.
 
\subsection{HIP 114569 and Cl VDBH 217}

The 5.5 M$_\odot$ star HIP 114569 was 25 pc away from the cluster VDBH 217, 7 Myrs ago. This is much more than the radius of the cluster, but no other cluster was found to be as close to the star. Furthermore, the cluster is slightly younger than the star at $40 \pm 10$ Myrs \citepads{2019AcA....69....1C,2017MNRAS.470.3937S}, and has only one star with known Gaia RV. Compared to the evolutionary age of the star and its flight time, BSS is the most probable scenario, which further makes this cluster unlikely as the parent. Therefore, it is likely that we have not yet identified a cluster for this star.

\subsection{HIP 105912 and Teutsch 145}
 
 The 10 M$_\odot$ star HIP 105912 had a close encounter 14 Myrs ago with the older cluster Teutsch 145 with a p-value of 0.66. The star is 20 Myrs old, leaving some time for binary evolution, and therefore BSS may be the origin here. However, firstly the radial velocity of the cluster relies on only one star, and secondly, age estimates for the cluster range from 200 Myrs \citeads{2017MNRAS.470.3937S} to more than 500 Myrs (\citeads{2013A&A...558A..53K}, \citeads{2016A&A...593A.116J}). Therefore, Teutsch 145 is unlikely to be the possible place of origin, if the cluster's radial velocity and age estimates are correct.
 
\subsection{BD-02 3766 and Kronberger 38}
The 10 M$_\odot$ star was near the cluster Kronberger 38 with a p-value of 0.54, almost 15 Myrs ago. This is also the estimated evolutionary age of the star, therefore making it a prime candidate for DES. Not much is known about the cluster itself, since it is in a region obscured by dust, and also only has one measured RV star.

 \subsection{BD-15 115 and Juchert 3/NGC 6755}

The 8 M$_\odot$ star B-15 115 is 32 Myrs old, and around the same time it was near 3 clusters: Juchert 3, NGC 6705, and NGC 6755 with p-values of 0.52, 0.33, and 0.01. However, NGC 6705 is older than 300 Myrs (\citeads{2021A&A...650A..67R},\citeads{Liu}) and therefore not a suitable choice for DES. On the other hand, not much is known about the cluster Juchert 3, except that is a slightly younger cluster \citepads{2013A&A...558A..53K}. However, Juchert 3 has only 1 RV star measured \citepads{OCDR2}. Finally, the lowest p-value cluster NGC 6755 is around 174 Myrs old \citepads{Liu}. Therefore, only Juchert 3 is a good candidate.

\section{The hyper-runaway candidate HIP 60350 and Cl Pismis 20}
\label{sect:hip60350}

\begin{table*}
\begin{center}
\caption{Possible Clusters as sites of origin of the Halo runaway stars using spatial coincidence.}
\hspace*{-0.5cm}
\renewcommand{\arraystretch}{1.3}
\begin{tabular}{ c c c c c c c c c c c} 
\hline
\hline
Star&Cluster Name& p-val & T$_f$ &T$_{rel}$& v$_\textrm{rel}$ & T$_c$&Flag$_{age}$&Suggested&References\\
&&&Myrs&Myrs&km\,s$^{-1}$&Myrs&&mechanism&&\\
\hline
HIP 60350&Cl Pismis 19&0.40&-14.3&-& 403&$1000\pm 230$&N&-&1,2,5\\
&Cl Pismis 20&0.43&-14.98&$9.90\pm0.13$& 376 &$<$50&Y&DES&1,2,6\\
&NGC 5617 & 0.002&-13.28 &-&378 &$199.5\pm 68.8$&N&-&1,2,5 \\
PG 1332+137\tablefootmark{a}&Sai 132&0.99&-20.7&-&272&500&N&-&1,2,7\\
PHL 159&Cl Basel 2&0.73&-24.9&-&137.4&$560\pm 70$&N&-&1,2,8\\
PB 5418&Cl Gulliver 43&0.67&-23.6&$25.72\pm1.45$&238&-&-&BSS&1,2\\
HIP 70275&NGC 1502&0.52&-11.2&$11.44\pm0.39$&260&$11.0\pm0.7$&Y&DES&1,3,9,10\\
HIP 11809&[FSR2007] 0883&0.05&-10.2&-&268.4&$700\pm42$&N&-&1,4,10\\
HIP 114569\tablefootmark{b}&Cl VDBH 217&-&-7&$7.52\pm1.21$&-&$40\pm10$&Y&BSS&1,2,11\\
HIP 105912&Teutsch 145&0.66&-13.9&-&202.3&500&N&-&1,2,7\\
BD-02 3766&Kronberger 38&0.54&-14.7&$16.18\pm1.08$&338.1&-&-&DES&1,2\\
BD-15 115&Juchert 3&0.52&-32&$30.54\pm0.87$&202.2&-&-&DES&1,2\\
&NGC 6755&0.016&-30.8&-&194.3&$174.0\pm10.4$\tablefootmark{c}&N&-&1,2,11,12\\
\hline
\end{tabular}
 \tablefoot{The columns show the p-values for EDR3 proper motions along with the expected time of flight T$_f$ in Myrs, and the relative velocities v$_\textrm{rel}$ in km\,s$^{-1}$. T$_c$ is the age of the clusters in Myrs. We use the column Flag$_{age}$ for a temporal origin based on the cluster age, with 'Y' meaning the cluster may be a possible site of origin, and 'N' meaning it cannot be a site since it is too old. \tablefoottext{a}{The star has a RUWE of 1.48 in EDR3. We therefore also ran the fit with DR2 data which has a RUWE of 0.99. The results did not change for the latter. },\tablefoottext{b}{Closest distance of encounter was 25 pc away.},\tablefoottext{c}{Age of 50 Myrs according to the Webda catalog.}}
 \tablebib{ Kinematic of clusters/associations: (1) \citetads{DR2OC}; (2)  RV \citetads{OCDR2};(3) \citetads{RAVEOC};(4) \citetads{clusterRV2019}; Age estimates of Clusters: (5) \citetads{2021MNRAS.500.4338A}; (6) \citetads{2018MNRAS.473..849D};(7) \citeads{2013A&A...557A..10N}; (8) \citetads{2010AstL...36...14G};(9) \citetads{2018AJ....155...91Y}; (10) \citetads{Liu}; (11) \citetads{2019AcA....69....1C}; (12) \citetads{2020A&A...644A.136A} }
\label{tab:haloidenfication}
\end{center}
\end{table*}

The star HIP 60350 was considered as a hyper-runaway candidate with a space velocity of about 400 km\,s$^{-1}$ by \citetads{andreasstar}. It was discussed by \citetads{2001A&A...369..530T} as having a possible origin in the cluster NGC 3603. However, due to the difference in the cluster age and the stellar age, the evidence for this origin was quite weak and was disproven by \citetads{andreasstar}. In their work, \citetads{andreasstar} presented a detailed kinematic and spectroscopic analysis of the star and suggested clusters which were close to the location of the star when it had crossed the disk of the Milky Way. However, a definitive place of origin is still lacking, although EDR3 data has constrained the stellar kinematics \citepads{andreasnew}. We adopt a radial velocity of v$_\textrm{rad}$=263.6 $\pm$ 0.5 km\,s$^{-1}$ as used by \citetads{andreasnew}, and use the proper motions according to \citetads{2021A&A...649A.124C}. 

We used our fitting procedure and were able to falsify all previously suggested origins in those clusters older than the stellar age (\citeads{andreasnew} place the star at a stellar age of 50 Myrs, much older than the previously suggested clusters). The only cluster left is the cluster NGC 5617, for which we get a p-value of 0.002 for a close encounter of less than 1 pc, which while sufficient might not be the best. Similar to section 4, we searched again for a new cluster using \textit{Gaia} EDR3 values, and found close encounters of the star with the clusters Cl Pismis 19 and Cl Pismis 20, with p-values of 0.40 and 0.43 (see Table \ref{tab:haloidenfication}). The first encounter happened 14.3 Myrs ago while the second was 14.98 Myrs ago, same as the time of disk crossing of 14.9 $\pm$ 1.0 suggested by \citetads{andreasstar}. Cl Pismis 20 is slightly younger than 50 Myrs \citepads{2018MNRAS.473..849D}, but may still be a suitable choice for the origin. \footnote{It must be mentioned that for the radial velocity of Cl Pismis 20, only 1 star was used by \citeads{OCDR2}, while 3 stars were used for Cl Pismis 19.} On the other hand Cl Pismis 19 is a much older cluster of around 1 Gyrs of age \citepads{2021MNRAS.500.4338A} and is therefore not a viable candidate.

\begin{figure}[htp]
\centering
\includegraphics[width=0.45\textwidth]{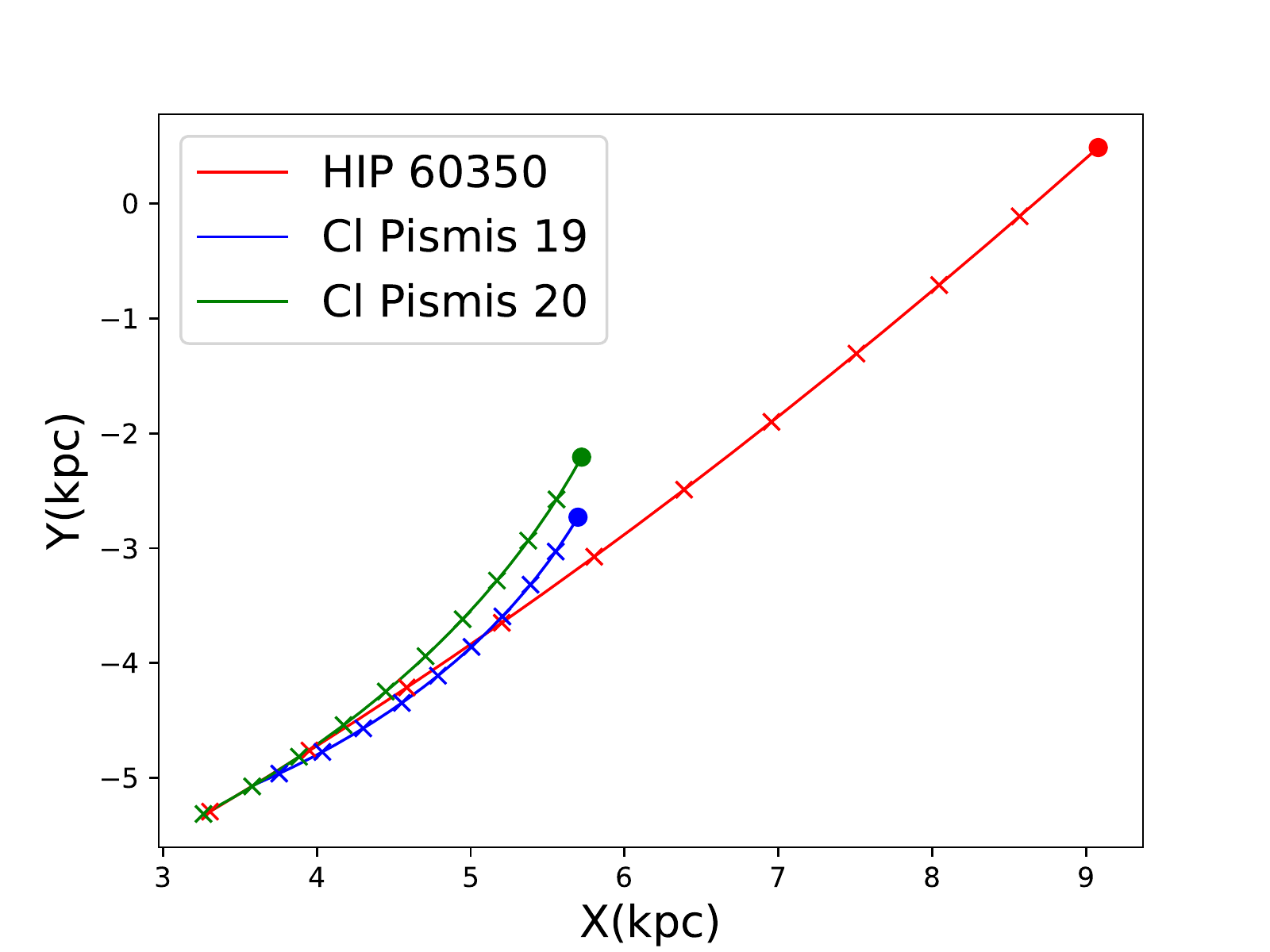}
\includegraphics[width=0.45\textwidth]{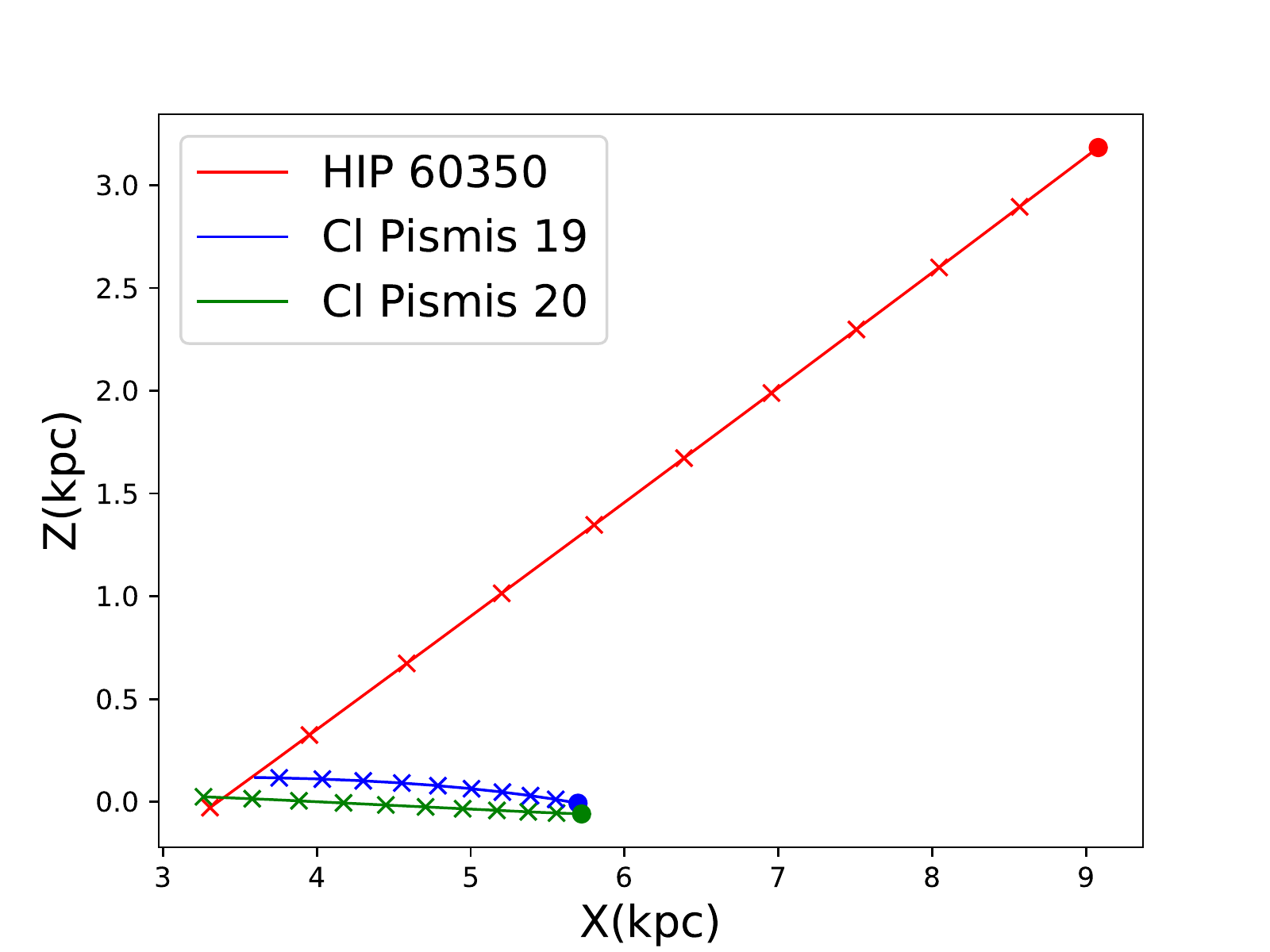}
\caption{The 3 dimensional orbit of the star HIP 60350 shown here in X-Y and X-Z Cartesian coordinates. The two cluster Cl Pismis 19 and Cl Pismis 20 are also plotted for 14.3 and 14.9 Myrs respectively. Crosses represent 1.5 Myrs in time while circles represent the present day positions. Trajectories are plotted using the revised halo potential of Model I from \citetads{andreassmass} in Galpy \citepads{bovy}.
}
\label{fig:orbithip60350}
\end{figure}

Since the stellar age is much older than the kinematic age, this star was possibly involved in a supernova explosion. According to \citetads{andreasstar}, one explanation could be the involvement of a high mass star which would be a Wolf-Rayet star of stellar type WN. Such stars have been known to be hosted by clusters, such as Cl Pismis 20 \citepads{2020MNRAS.495.1209R}. The mass required for the Wolf-Rayet star would be greater than 15 M$_\odot$. Such a star would now be either a neutron star or a black hole, 
rendering it difficult 
to find the original companion of HIP 60350. Fig.\ref{fig:orbithip60350} shows the orbit of the star and the clusters, illustrating the high latitude nature of the star and how it may have been ejected from the disk. The results for HIP 60350 along with all the other halo runaways are summarized in Table \ref{tab:haloidenfication}.

\section{Summary and conclusion}\label{sect:summary}

The excellent quality of \textit{Gaia} astrometry opens up the path to study the kinematics of young runaway stars to such a high precision that the place of origin in open stellar clusters can be identified uniquely even when it is a few kiloparsec away. Moreover, hundreds of new clusters have been discovered utilizing \textit{Gaia} data.  

To cope with the quality and growth of observational data, we developed an efficient minimization method to calculate whether two or more objects may come from the same place. Our method is based on a fitting procedure which fits the trajectories of objects by minimization of the distance between them taking into account the significance of the trajectories by means of the statistical p value. The $2\sigma$ limit for 8 parameters is set, and any p-value below this level allows us to reject the null hypothesis and conclude that the two or more objects do not come from the same place with $2\sigma$ certainty. We test our procedure first on the classical runaways AE Aurigae and $\mu$ Columbae thought to have originated in the ONC 2.5 Myrs ago. Furthermore we confirm the results of \citeads{2020MNRAS.495.3104S} and \citeads{Farias_orion} for stars thought to originate in the ONC, except for three stars that we found to have closest approaches of more than 40 pc from the ONC center.

So far, the most sophisticated study of cluster runaways was carried out by \citetads{hoog} based on \textit{Hipparcos} data.
Using new data from \textit{Gaia} DR2 and EDR3, and \textit{Hipparcos} new reduction, along with new radial velocities, we run our procedure on the runaway sample of \citeads{hoog}. We confirm the previous results for 53 Ari, AE Aurigae, $\psi$ Per, HIP 22061, and HIP 38518. Among the rest of the sample we find new candidate clusters as sites of origin using new \textit{Gaia} data for most of them. In many cases the reason for the new identifications can be traced back to the radial velocities used in the original study. Follow-up spectroscopy had shown that some stars are radial velocity variable binaries. We used the systemic velocities, which had a strong impact on the kinematic results. Only half of the sample is confirmed to be runaway stars, while the space velocities of the second half are so low ($<$30 km\,s$^{-1}$), that they do not qualify as runaways, but have to be considered walkaway stars.
In particular we show that the classical runaways AE Aurigae and $\mu$ Columbae might not have originated together, with $\mu$ Columbae having an earlier ejection from Collinder 69, a cluster near the ONC. Furthermore, the binary system $\iota$ Ori is found to not have had a close encounter with either AE Aurigae or the ONC.
We use cluster age and flight time of the stars to investigate whether the ejection was due to a binary supernova or due to a dynamical ejection.

We use our procedure on a list of runaway B stars in the halo compiled from \citetads{2011MNRAS.411.2596S} using results from recent analyzes by \citetads{Markus:Thesis:2021}. We are able to find candidate clusters for four stars where these stars and derive their ejection velocities to lie in the range 160--410 km\,s$^{-1}$\citepads{Markus:Thesis:2021} considerably larger than for the \citetads{hoog} sample. Three other star-cluster pairs may have a common origin, but due to nonavailability of ages for these clusters, we are unable to confirm this. We also find that in some cases the clusters may have been at the same places as the stars in the past, but the cluster ages rule out any common origin.

Finally, the hyper-runaway B-star HIP 60350 was revisited. The recent \textit{Gaia EDR3} based  kinematic analysis by \citetads{andreasnew} already dismissed the star as hyper-runaway star because its Galactic restframe velocity is  lower than the local escape velocity. All of the previously proposed parent clusters had to be dismissed, and based on both spatial and temporal coincidence, Pismis 20 turn out to be the only good candidate. The ejection velocity of 338 km\,s$^{-1}$ is lower than previously derived but is similar to that of the Galactic Halo runaway B stars.  

While radial velocities of the runaway stars are very well known, this is not the case for many open clusters, whose radial velocities are based on very few stars, occasionally on one star, only. This is especially true for open clusters more than 1 kpc away from the Sun. Additional radial velocity surveys of open clusters are needed to derive accurate cluster radial velocities, which would further allow their kinematics to be constrained better.

\begin{acknowledgements}

A.B., A.I. and U.H. acknowledge funding by the Deutsche Forschungsgemeinschaft under grants HE 1356/70-1 and IR 190/1-1. We would like to thank the anonymous referee for their insightful comments.
This work has made use of data from the European Space Agency (ESA) mission
{\it Gaia} (\url{https://www.cosmos.esa.int/gaia}), processed by the {\it Gaia}
Data Processing and Analysis Consortium (DPAC,
\url{https://www.cosmos.esa.int/web/gaia/dpac/consortium}). Funding for the DPAC
has been provided by national institutions, in particular the institutions
participating in the {\it Gaia} Multilateral Agreement. This research has made use
 of the SIMBAD database, operated at CDS, the VizieR catalog access tool, CDS,
 Strasbourg, France (DOI : 10.26093/cds/vizier), and TOPCAT table processing software \citepads{2005ASPC..347...29T}. 

\end{acknowledgements}

{
\footnotesize
\bibliography{main}
}
\medskip

\appendix

\section{Kinematic parameters for runaway stars, 
and open clusters used in this study}\label{sect:appendix:A}

Table \ref{tab:hypervelparams} list the astrometric data from \textit{Hipparcos} and \textit{Gaia} for the sample of \citetads{hoog} (see Sect. \ref{sect:hoogerwerf}), while Table \ref{tab:clusterparams} compiles the data for the relevant open cluster. In Table \ref{tab:hyperrun} the astrometric data for the runaway B stars are listed (see Sect. \ref{sect:high_vel} and Sect. \ref{sect:hip60350}).

\pgfplotstableread[col sep=comma]{tables/Hypervel_Hooglist.csv}\tableaf
\begin{table*}
\centering
\caption{Parameters of runaways from the \citetads{hoog} sample.  \textit{Hipparcos}, \textit{Gaia} DR2  \citepads{2018A&A...616A...2L}  and EDR3 \citepads{2021A&A...649A...2L} astrometric data are given.
\label{tab:hypervelparams}}
\renewcommand{\arraystretch}{1.5}
\resizebox{0.9\textwidth}{!}{
\pgfplotstabletypeset[
columns={hip_number_1,plx_hip,e_plx_hip,pmra_hip,e_pmra_hip,pmde_hip,e_pmde_hip,F2,plx_dr2,e_plx,pmra,e_pmra,pmde,e_pmde,ruwe_dr2,parallax_dr3,parallax_error_dr3,pmra_dr3,pmra_error_dr3,pmdec_dr3,pmdec_error_dr3,ruwe_dr3,rv,rv_error,V},
column type=c,
string type,
every head row/.style={before row={\hline\hline&\multicolumn{7}{c}{\textit{Hipparcos}}&\multicolumn{7}{c}{DR2}&\multicolumn{7}{c}{EDR3}\\},after row=\hline,},
every last row/.append style={after row={\hline} },
every first column/.style={column type/.add={}{}},
every last column/.style={column type/.add={}{}},
columns/hip_number_1/.style={column name=HIP,string replace*={_}{\textunderscore}},
columns/plx_hip/.style={column name=$\varpi$,column type=c,numeric type,fixed,precision=2},
columns/e_plx_hip/.style={column name=$e_\varpi$,column type=c,numeric type,fixed,precision=2},
columns/pmra_hip/.style={column name=$\mu_{\alpha*}$,column type=c,numeric type,fixed,precision=2},
columns/e_pmra_hip/.style={column name=e$\mu_{\alpha*}$,column type=c,numeric type,fixed,precision=2},
columns/pmde_hip/.style={column name=$\mu_{\delta}$,column type=c,numeric type,fixed,precision=2},
columns/e_pmde_hip/.style={column name=e$\mu_{\delta}$,column type=c,numeric type,fixed,precision=2},
columns/F2/.style={column name=F2,column type=c,numeric type,fixed,precision=2},
columns/plx_dr2/.style={column name=$\varpi$,column type=c,numeric type,fixed,precision=2},
columns/e_plx/.style={column name=$e_\varpi$,column type=c,numeric type,fixed,precision=2},
columns/pmra/.style={column name=$\mu_{\alpha*}$,column type=c,numeric type,fixed,precision=2},
columns/e_pmra/.style={column name=e$\mu_{\alpha*}$,column type=c,numeric type,fixed,precision=2},
columns/pmde/.style={column name=$\mu_{\delta}$,column type=c,numeric type,fixed,precision=2},
columns/e_pmde/.style={column name=e$\mu_{\delta}$,column type=c,numeric type,fixed,precision=2},
columns/ruwe_dr2/.style={column name=RUWE(DR2),column type=c,numeric type,fixed,precision=2},
columns/parallax_dr3/.style={column name=$\varpi$,column type=c,numeric type,fixed,precision=2},
columns/parallax_error_dr3/.style={column name=e$\varpi$,column type=c,numeric type,fixed,precision=2},
columns/pmra_dr3/.style={column name=$\mu_{\alpha*}$,column type=c,numeric type,fixed,precision=2},
columns/pmra_error_dr3/.style={column name=e$\mu_{\alpha*}$,column type=c,numeric type,fixed,precision=2},
columns/pmdec_dr3/.style={column name=$\mu_{\delta}$,column type=c,numeric type,fixed,precision=2},
columns/pmdec_error_dr3/.style={column name=e$\mu_{\delta}$,column type=c,numeric type,fixed,precision=2},
columns/ruwe_dr3/.style={column name=RUWE(EDR3),column type=c,numeric type,fixed,precision=2},
columns/rv/.style={column name=RV,column type=c,numeric type,fixed,precision=2},
columns/rv_error/.style={column name=e$_{RV}$,column type=c,numeric type,fixed,precision=2},
columns/V/.style={column name=V mag,column type=c,numeric type,fixed,precision=2},
]{\tableaf}
}
\tablefoot{\textit{Hipparcos} second reduction is used here. HIP 26241 is $\iota$ Ori which has no entry in  \textit{Gaia}  DR2.}
\end{table*}

\pgfplotstableread[col sep=comma]{tables/cluster_parameters.csv}\tableaf
\begin{table*}[ht]
\centering
\caption{Kinematic parameters of stellar clusters. 
\label{tab:clusterparams}}
\renewcommand{\arraystretch}{1.1}
\resizebox{0.9\textwidth}{!}{
\pgfplotstabletypeset[
columns={SimbadName_1,Plx,e_Plx,pmRA,e_pmRA,pmDE,e_pmDE,RV,s_RV,N,NStars},
column type=c,
string type,
every head row/.style={before row=\hline\hline,after row=\hline,},
every last row/.append style={after row={\hline} },
every first column/.style={column type/.add={}{}},
every last column/.style={column type/.add={}{}},
columns/SimbadName_1/.style={column name=Name,string replace*={_}{\textunderscore}},
columns/Plx/.style={column name=$\varpi$,column type=c,numeric type,fixed,precision=2},
columns/e_Plx/.style={column name=$e_\varpi$,column type=c,numeric type,fixed,precision=3},
columns/pmRA/.style={column name=$\mu_{\alpha*}$,column type=c,numeric type,fixed,precision=2},
columns/e_pmRA/.style={column name=e$\mu_{\alpha*}$,column type=c,numeric type,fixed,precision=2},
columns/pmDE/.style={column name=$\mu_{\delta}$,column type=c,numeric type,fixed,precision=2},
columns/e_pmDE/.style={column name=e$\mu_{\delta}$,column type=c,numeric type,fixed,precision=2},
columns/RV/.style={column name=RV,column type=c,numeric type,fixed,precision=2},
columns/s_RV/.style={column name=eRV,column type=c,numeric type,fixed,precision=2},
columns/N/.style={column name=N$_\text{PM}$,column type=c,numeric type,fixed},
columns/NStars/.style={column name=N$_\text{RV}$,column type=c,numeric type,fixed,precision=2},
]{\tableaf}
}
\tablefoot{The parallaxes, the proper motions, and the radial velocities along with their respective errors are taken from literature (see text). N$_\text{PM}$ and N$_\text{RV}$ are the number of stars used for the proper motions and radial velocity calculations respectively. N$_\text{RV}$ with a value of 0 are the values taken from \citeads{RAVEOC}, where the number of stars is not mentioned. The clusters are shown in the descending order of their parallaxes, which explains the decrease in number of radial velocity stars for clusters farther away.}
\end{table*}

\setlength{\tabcolsep}{0.40em}
\begin{table*}
\centering
\caption{Kinematic parameters of B type runaways in the Galactic halo taken from \citeads{Markus:Thesis:2021}}
\begin{tabular}{lcccccccc}
 \hline\hline
Object & $\mu_\alpha\cos{\delta}$ & $\mu_\delta$ & $\varpi$ & RUWE& $v_\mathrm{rad}$&$d_\mathrm{spec}$&$t_\mathrm{evol}$&$t_\mathrm{flight}$\\
& (mas/yr) & (mas/yr) & (mas) & &(km\,s$^{-1}$)&Kpc&Myr&Myr\\ \hline
 BD $-$15 115 & $4.29\,\pm\,0.06$ & $-1.03\,\pm\,0.06$ & $0.18\,\pm\,0.05$ & $0.77$&  $90.5^{+1.3}_{-1.3}$&$4.57^{+0.22}_{-0.22}$&$32.4^{+1.1}_{-1.1}$&$31.0^{+0.8}_{-0.9}$ \\
 BD $-$2 3766 & $0.69\,\pm\,0.06$ & $17.25\,\pm\,0.04$ & $0.22\,\pm\,0.05$ & $1.08$& $25.2^{+1.0}_{-1.0}$ &$3.46^{+0.17}_{-0.17}$&$16.7^{+0.9}_{-1.2}$&$14.7^{+0.1}_{-0.1}$\\
 HD 151397 & $-9.26\,\pm\,0.03$ & $-0.39\,\pm\,0.02$ & $0.52\,\pm\,0.02$ & $0.81$& $149.6^{+0.6}_{-0.6}$&$1.66^{+0.08}_{-0.08}$&$1.3^{+1.3}_{-1.0}$&$1.57^{+0.005}_{-0.005}$\\
 HIP 105912 & $11.04\,\pm\,0.05$ & $-11.61\,\pm\,0.04$ & $0.46\,\pm\,0.05$ & $0.98$ & $2.5^{+5.4}_{-5.4}$&$2.21^{+0.12}_{-0.11}$&$20.0^{+0.8}_{-0.8}$&$14.0^{+0.4}_{-0.4}$ \\
 HIP 114569 & $45.76\,\pm\,0.05$ & $32.22\,\pm\,0.04$ & $0.73\,\pm\,0.05$ & $1.20$& $99.7^{+1.5}_{-1.5}$&$1.64^{+0.08}_{-0.08}$&$52.8^{+2.6}_{-3.8}$&$7.5^{+0.2}_{-0.2}$\\
 HIP 11809 & $-20.67\,\pm\,0.04$ & $-12.05\,\pm\,0.04$ & $0.53\,\pm\,0.04$ & $1.05$& $2.5^{+9.0}_{-9.0}$&$2.25^{+0.11}_{-0.11}$&$103^{+19}_{-21}$&$11.3^{+0.5}_{-0.5}$ \\
 HIP 56322 & $2.91\,\pm\,0.07$ & $12.03\,\pm\,0.05$ & $0.34\,\pm\,0.06$ & $1.12$& $260.3^{+1.3}_{-1.3}$&$3.10^{+0.15}_{-0.15}$&$12.5^{+1.7}_{-2.0}$&$8.3^{+0.3}_{-0.3}$ \\
 HIP 60350 & $-13.30\,\pm\,0.04$ & $15.03\,\pm\,0.05$ & $0.28\,\pm\,0.05$ & $1.29$& $263.6^{+0.4}_{-0.5}$&$3.35^{+0.16}_{-0.16}$&$55^{+5}_{-6}$&$15.3^{+0.9}_{-0.8}$ \\
 HIP 70275 & $3.25\,\pm\,0.06$ & $-10.32\,\pm\,0.05$ & $0.39\,\pm\,0.06$ & $1.38$& $242.3^{+0.2}_{-0.2}$ &$2.30^{+0.11}_{-0.11}$&$17.6^{+1.2}_{-1.7}$&$13.3^{+0.9}_{-0.9}$\\
 PB 5418 & $1.67\,\pm\,0.06$ & $-2.87\,\pm\,0.03$ & $0.17\,\pm\,0.04$ & $0.99$& $146.6^{+2.6}_{-2.6}$ &$6.10^{+0.29}_{-0.29}$&$42.4^{+1.9}_{-1.9}$&$23.3^{+0.7}_{-0.7}$\\
 PG 1332$+$137 & $-6.57\,\pm\,0.06$ & $-8.46\,\pm\,0.03$ & $0.16\,\pm\,0.05$ & $1.48$& $160.5^{+1.7}_{-1.7}$&$3.87^{+0.18}_{-0.18}$ &$46^{+5}_{-6}$&$20.6^{+0.7}_{-1.0}$\\
 PHL 159 & $-6.15\,\pm\,0.05$ & $-9.63\,\pm\,0.04$ & $0.39\,\pm\,0.05$ & $1.11$& $87.8^{+0.6}_{-0.6}$&$3.48^{+0.17}_{-0.17}$&$21.2^{+1.0}_{-1.5}$&$24.0^{+0.7}_{-0.8}$ \\
\hline\end{tabular}
\tablefoot{$t_\mathrm{flight}$ denotes the flight time from the disk of the milky way according to \citetads{Markus:Thesis:2021}}
\label{tab:hyperrun}
\end{table*}





\section{Testing the p-value method for the Hoogerwerf+ sample and mock data}\label{sect:appendix:B}
For the purpose of testing our methods as well as to see how any predictions deal with uncertainties we created some mock data. For this we recreated an interaction between AE Aurigae and $\mu$ Columbae in the present time. We took the present position of the Orion Nebula cluster (ONC, Trapezium cluster) and assumed the two stars to be in the center of the ONC. We provided the same parallax to both the stars assuming they come from the same point in space. Then we applied our orbit calculator on the two stars using their present day velocities (this may seem nonphysical at first but the initial velocities are not as constrained in the interaction as the three-dimensional position). We ran the two orbits for 2.5 Myrs (approximately the flight time of present day AE Aurigae and Mu Col.) and noted the positions and velocities of the two stars now. The common starting spatial parameters used for the two stars were:\\ 
$\alpha$: 5h\,35m\,16s; $\delta$ : -5$^\circ$\,23'\,40''; $\varpi$: 2.50 mas *

\begin{figure*}[h]
\hspace{-0.5cm}
\centering
\includegraphics[width=1.0\textwidth]{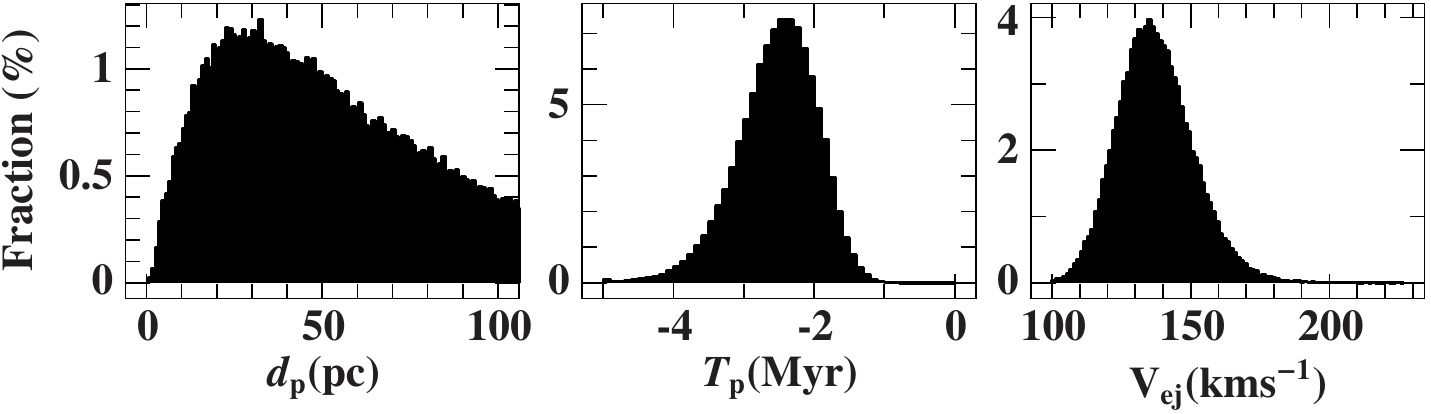}
\includegraphics[width=1.0\textwidth]{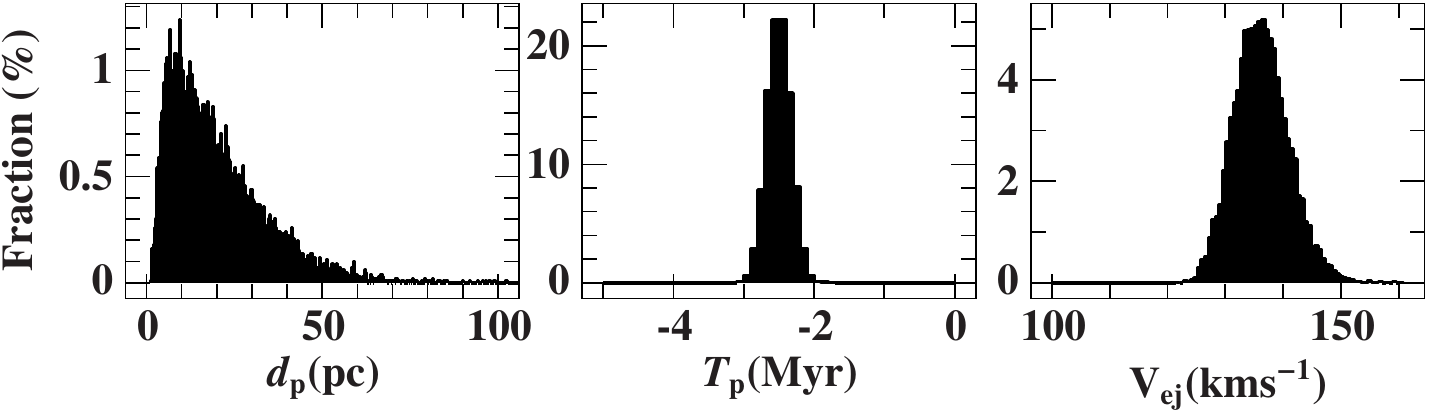}
\caption{Histograms for simulated back-traced mock orbits of two stars, with 10\% errors (top) and EDR3 based relative errors on AE Aurigae and $\mu$ Col (bottom). The left most panel shows the fraction of distances of closest encounters between the two stars. The middle panel is the histogram of the times for the closest encounters of the two stars, and the right most panel plots the relative {ejection} velocity of the stars with respect to each other.}
\label{fig:mock1}
\end{figure*}

Therefore, we ended up with two different trajectories and used the end points of these trajectories as the values for our mock data. 
To simulate \textit{Hipparcos} and \textit{Gaia} precision we add to all the final velocities and parallaxe in the first simulation 10\% errors, which is typical for \textit{Hipparcos} precision. In a second simulation we
the same relative errors as in EDR3 catalog.  
Finally, we trace back the two stars in time again for both adopted precisions. To check whether our mock data are consistent, we apply the simulation technique by computing 100000 total backward trajectories and plot histograms shown in Fig. \ref{fig:mock1}. The results of the simulation using Gaia precision (lower panel of Fig. \ref{fig:mock1}) are close to that derived from the new observations shown in Fig. \ref{fig:hist}. The exercise demonstrates the great improvement to be achieved from high precision measurements.

Our fitting procedure found a p-value of 0.85, even in the simulation with 10\% errors, close to the values found for real data of \textit{Hipparcos}. 
 
\begin{figure*}[h]
\hspace{-0.5cm}
\centering
\includegraphics[width=1.0\textwidth]{figures/histogramsaeaur_mucol_edr3_mock_nocorr_all.pdf}
\includegraphics[width=1.0\textwidth]{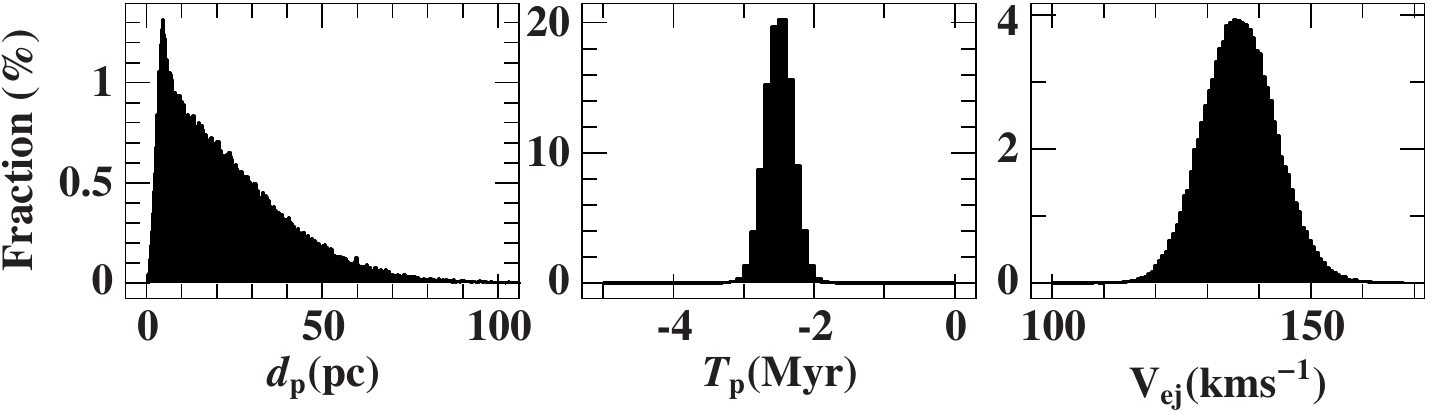}
\caption{Histograms for simulated back-traced mock orbits of two stars with
EDR3 based relative errors on AE Aurigae and $\mu$ Col (bottom), when a correlation factor of 0.9 is added to all the three major parameters (top). The left most panel shows the fraction of distances of closest encounters between the two stars. The middle panel is the histogram of the times for the closest encounters of the two stars, and the right most panel plots the relative {ejection} velocity of the stars with respect to each other.}
\label{fig:mockcor}
\end{figure*}

Our MC simulations allow us to include astrometric correlations, which are available for Gaia DR2 \citepads{2018A&A...616A...2L} and EDR3 \citepads{2021A&A...649A...2L} data.
To test for the effects of such correlations, we added correlation factors of 0.9 between the parallax, proper motion in right ascension, and proper motion in declination. The histograms in Fig. \ref{fig:mockcor} show the effect of these correlations. For EDR3 based errors, the errors are small enough that the correlations do not lead to significant changes in the inference of the trajectories. A similar result is found when the correlation is 0.5.

Finally, we also ran our p-value based distance minimization procedure for the runaway stars and clusters to identify their places of origin suggested by \citetads{hoog} using the original radial velocities and proper motions in most cases as a proof of test for our method to compare our results with the published one. 
For clusters we utilize values from the Simbad database, based on new \textit{Gaia} DR2 kinematics where available. Table \ref{tab:table1} shows our results. 

\begin{center}
\begin{table*}[ht]
\centering
\renewcommand{\arraystretch}{1.3}
\caption{ P values for the list of stars in the sample of \citetads{hoog}. The column p-val lists the p-values and the column T$_f$ lists the best fit times of flight for the \textit{Hipparcos} astrometry.}
\begin{tabular}{c c c c c c} 
\hline
\hline
Star Name & HIP Number & Star/Place of Origin &p-val (Hip)&T$_f$&Reference\\
    &                  6                      &      &  Myrs &\\
\hline
$\nu$ And&3881&Lac OB1&None&-&1\\
53 Ari&14514&Orion OB1&0.39&-5.42&2 \\

$\xi$ Per  &18614&IC 0348&0.976&-1.09&3\\
 &&IC 0348 DR2 &None&-&  4\\
&& Per OB2 Ass.&None&-& 3\\

HD 30112& 22061 &HIP 29678&0.751&-1.16&3\\

AE Aur&24575& $\mu$ Col. & 0.87&-2.16&\\ 
AE Aur&& Orion & 0.73& -2.53&5 \\

$\mu$ Col.& 27204&Orion & 0.75& -2.37& 5\\

b Pup& 38455&Collinder 135&None&-& 4,6\\

J Pup& 38518  &Vel OB2&0.54&-9.49& 3\\

$\zeta$ Pup &39429&Trumpler 10&0.63&-2.61&4, 6\\
&&Trumpler 10&0.94\tablefootmark{a}&-2.44& 7\\

HD 73105& 42038  &IC 2391&0.40&-7.70& 3\\
  &&IC 2391&0.54&-7.95& 4\\
&&Sco-Cen OB2/UCL&$<2\sigma$&-&1,8\\

* L Vel& 46950 &IC 2602&0.63&-9.64& 3\\
 &&IC 2602&0.83&-9.92& 4\\
 
HD 88661& 49934  &IC 2602&0.05&-9.02& 3\\
 &&IC 2602&0&-9.32&4\\
  &&IC 2391&None&-& 3\\
 & &IC 2391&0.35&-9.74& 4\\

HD 102776 & 57669 & IC 2602&None&-& 3\\
 &&IC 2602&None&-& 4\\

V716 Cen &69491&UCL&0.93&-3.0,5 pc&8\\

HD 137387&76013&LCC&None&-&8\\

$\zeta$ Oph & 81377&PSR B1929+10\tablefootmark{b}&0.7835&-0.829&\\
 & &PSR B1929+10\tablefootmark{c}&None&-&\\

 && PSR B1706-16\tablefootmark{d}&0.41& -1.33&\\

 & &PSR B1706-16\tablefootmark{e}&$<2\sigma$&-1.47&\\

HD 152478& 82868 &IC 2602&0.11&-11.8&  3 \\
 &&IC 2602&0.19&-11.07& 4\\

HD 172488& 91599 &IC 0348 DR2 &0.23&-8.79& 4\\
&& Per OB2 Ass.&0&-9.48&3\\
&&CL Melotte 20/Per OB3 &0.63&-7.34& 7 \\
&& &0.83&-7.40& 4\\

$\lambda$ Cep  &109556&Cep OB3&None&-&3,9\\

  &&Cep OB3b&0.17&-8.22&5\\
& &Tr37&0.02&-10& 10,11\\
 & &NGC 7160&None&-&12\\
 &&NGC 7160&0.003&-11.42&5\\
\hline
\end{tabular}
\tablebib{(1) \citetads{2009MNRAS.400..518M};(2) \citetads{2017MNRAS.472.3887M};(3) Simbad values;(4) \citetads{DR2OC};(5) \citetads{2019ApJ...870...32K};(6) RV \citetads{OCDR2}; (7) \citetads{GaiaHR}; (8) \citetads{1999AJ....117..354D};(9) $\varpi$ adopted \citepads{2018MNRAS.473..849D};(10) $\varpi$, PM \citetads{DR2OC2};(11) RV \citetads{RAVEOC}; }
\tablefoot{\tablefoottext{a}{1.4 pc},\tablefoottext{b}{Old values for RV and PSR},\tablefoottext{c}{New values for RV and PSR},
\tablefoottext{d}{New RV, PSR PM from \citetads{1997MNRAS.286...81F}},\tablefoottext{e}{New values for RV, PSR PM from \citetads{2019MNRAS.484.3691J}}}
 \label{tab:table1}
\end{table*}
\end{center}

\begin{center}
\begin{table*}[ht]
\caption{ P values for the previously known single and binary stars in the sample of \citetads{hoog} with \textit{Gaia}   DR2 \citepads{2018A&A...616A...2L} and EDR3 \citepads{2021A&A...649A...2L} astrometries. }
\centering
\renewcommand{\arraystretch}{1.3}
\begin{tabular}{ c c c c c c c c } 
\hline
\hline
Star Name&HIP Number&Star/Place of Origin &p-val (DR2)&T$_f$(DR2)&p-val (EDR3)&T$_f$(EDR3)&Reference\\
\hline
$\nu$ And&3881&Lac OB1&None&&None&&1\\
53 Ari&14514&Orion OB1&0.58& -5.20&0.54&-5.24&2 \\

$\xi$ Per  &18614&IC 0348&0.641&-1.29& None $<$ 0.0012&&3\\
 &&IC 0348 DR2 &0.07&6.75&None&&  4\\
&& Per OB2 Ass.&0.295&-6.44&None&& 3\\

HD 30112& 22061 &HIP 29678&0.851&-1.10&0.994&-1.08&3\\

AE Aur&24575& $\mu$ Col. & 0.1&&0.7&&\\ 
AE Aur&& Orion & 0.77&-2.58&None&& 5 \\

$\mu$ Col.& 27204&Orion &0.03&-2.35&0.73&-2.52& 5\\

J Pup& 38518  &Vel OB2&0.75&-4.34&0.0003&-5.30& 3\\

$\zeta$ Pup &39429&Trumpler 10&----&&-----&&4, 6\\
&&Trumpler 10&----&&-----& &7\\

*
HD 88661& 49934  &IC 2602&None&&None&&  3\\
 &&IC 2602&None&&None&&4\\
  &&IC 2391&None&&None&& 3\\
 & &IC 2391&None&&None&&  4\\

$\zeta$ Oph & 81377&PSR B1929+10\tablefootmark{a}&0.594&-0.961&0.682&-0.776&\\
 & &PSR B1929+10\tablefootmark{b}&None&&None&&\\

 && PSR B1706-16\tablefootmark{c}&0.90&-1.33&0.96&-1.38&\\

 & &PSR B1706-16\tablefootmark{d}&None&&$<2\sigma$&-1.98&\\

HD 152478& 82868 &IC 2602&None&&None&&  3 \\
 &&IC 2602&None&&None&& 4\\

HD 172488& 91599 &IC 0348 DR2 &None&&None&&  4\\
&& Per OB2 Ass.&None&&None&&3\\
&&CL Melotte 20/Per OB3 &None&&None&& 7 \\
&& &None&&None&& 4\\

$\lambda$Cep  &109556&Cep OB3&None&&None&&3,8\\

  &&Cep OB3b&None&&None&&5\\
& &Tr37&None&&None&& 9,10\\
 & &NGC 7160&None&&None&&11\\
 &&NGC 7160&None&&None&&3\\
\hline
\end{tabular}
\tablebib{(1) \citetads{2009MNRAS.400..518M}; (2) \citetads{2017MNRAS.472.3887M};(3) Simbad; (4) \citetads{DR2OC};(5) \citetads{2019ApJ...870...32K};(6) RV from \citetads{OCDR2};(7) \citepads{GaiaHR};(8) $\varpi$ adopted \citetads{2018MNRAS.473..849D};(9) $\varpi$, PM \citepads{DR2OC2};(10) RV \citepads{RAVEOC};(11) $\varpi$, PM, RV from \citetads{2019ApJ...870...32K}; }
 \tablefoot{The columns with p-val report the p-values while T$_f$ is the best fit flight time.\tablefoottext{a}{Old values for RV and PSR}\tablefoottext{b}{New values for RV and PSR}
\tablefoottext{c}{New RV, PSR PM from \citetads{1997MNRAS.286...81F}}\tablefoottext{d}{New values for RV, PSR PM from \citetads{2019MNRAS.484.3691J}}}
 \label{tab:oldtars}
\end{table*}
\end{center}

\end{document}